\documentclass[twocolumn,reprint,amsmath,amssymb,aps,floatfix,groupedaddress,superscriptaddress,nofootinbib,preprintnumbers]{revtex4-1}
\usepackage{float}
\usepackage{amsmath}
\usepackage[pdftex]{hyperref}
\usepackage{graphicx,color}
\usepackage{dcolumn}
\usepackage{bm}
\usepackage{here}
\usepackage{comment}
\usepackage[paperwidth=210mm,paperheight=297mm,centering,hmargin=1.5cm,vmargin=2.0cm]{geometry}
\usepackage{tabularx}
\usepackage{multirow}

\allowdisplaybreaks[1]

\makeatletter
\newsavebox{\@brx}
\newcommand{\llangle}[1][]{\savebox{\@brx}{\(\m@th{#1\langle}\)}%
  \mathopen{\copy\@brx\mkern2mu\kern-0.9\wd\@brx\usebox{\@brx}}}
\newcommand{\rrangle}[1][]{\savebox{\@brx}{\(\m@th{#1\rangle}\)}%
  \mathclose{\copy\@brx\mkern2mu\kern-0.9\wd\@brx\usebox{\@brx}}}
\makeatother

\begin{document}

\title{ Measurement of the sixth-order cumulant of net-proton multiplicity distributions \\in Au+Au collisions at $\sqrt{s_{\rm NN}}=$~27, 54.4, and 200~GeV at RHIC } 

\affiliation{Abilene Christian University, Abilene, Texas   79699}
\affiliation{AGH University of Science and Technology, FPACS, Cracow 30-059, Poland}
\affiliation{Alikhanov Institute for Theoretical and Experimental Physics NRC "Kurchatov Institute", Moscow 117218, Russia}
\affiliation{Argonne National Laboratory, Argonne, Illinois 60439}
\affiliation{American University of Cairo, New Cairo 11835, New Cairo, Egypt}
\affiliation{Brookhaven National Laboratory, Upton, New York 11973}
\affiliation{University of California, Berkeley, California 94720}
\affiliation{University of California, Davis, California 95616}
\affiliation{University of California, Los Angeles, California 90095}
\affiliation{University of California, Riverside, California 92521}
\affiliation{Central China Normal University, Wuhan, Hubei 430079 }
\affiliation{University of Illinois at Chicago, Chicago, Illinois 60607}
\affiliation{Creighton University, Omaha, Nebraska 68178}
\affiliation{Czech Technical University in Prague, FNSPE, Prague 115 19, Czech Republic}
\affiliation{Technische Universit\"at Darmstadt, Darmstadt 64289, Germany}
\affiliation{ELTE E\"otv\"os Lor\'and University, Budapest, Hungary H-1117}
\affiliation{Frankfurt Institute for Advanced Studies FIAS, Frankfurt 60438, Germany}
\affiliation{Fudan University, Shanghai, 200433 }
\affiliation{University of Heidelberg, Heidelberg 69120, Germany }
\affiliation{University of Houston, Houston, Texas 77204}
\affiliation{Huzhou University, Huzhou, Zhejiang  313000}
\affiliation{Indian Institute of Science Education and Research (IISER), Berhampur 760010 , India}
\affiliation{Indian Institute of Science Education and Research (IISER) Tirupati, Tirupati 517507, India}
\affiliation{Indian Institute Technology, Patna, Bihar 801106, India}
\affiliation{Indiana University, Bloomington, Indiana 47408}
\affiliation{Institute of Modern Physics, Chinese Academy of Sciences, Lanzhou, Gansu 730000 }
\affiliation{University of Jammu, Jammu 180001, India}
\affiliation{Joint Institute for Nuclear Research, Dubna 141 980, Russia}
\affiliation{Kent State University, Kent, Ohio 44242}
\affiliation{University of Kentucky, Lexington, Kentucky 40506-0055}
\affiliation{Lawrence Berkeley National Laboratory, Berkeley, California 94720}
\affiliation{Lehigh University, Bethlehem, Pennsylvania 18015}
\affiliation{Max-Planck-Institut f\"ur Physik, Munich 80805, Germany}
\affiliation{Michigan State University, East Lansing, Michigan 48824}
\affiliation{National Research Nuclear University MEPhI, Moscow 115409, Russia}
\affiliation{National Institute of Science Education and Research, HBNI, Jatni 752050, India}
\affiliation{National Cheng Kung University, Tainan 70101 }
\affiliation{Nuclear Physics Institute of the CAS, Rez 250 68, Czech Republic}
\affiliation{Ohio State University, Columbus, Ohio 43210}
\affiliation{Institute of Nuclear Physics PAN, Cracow 31-342, Poland}
\affiliation{Panjab University, Chandigarh 160014, India}
\affiliation{Pennsylvania State University, University Park, Pennsylvania 16802}
\affiliation{NRC "Kurchatov Institute", Institute of High Energy Physics, Protvino 142281, Russia}
\affiliation{Purdue University, West Lafayette, Indiana 47907}
\affiliation{Rice University, Houston, Texas 77251}
\affiliation{Rutgers University, Piscataway, New Jersey 08854}
\affiliation{Universidade de S\~ao Paulo, S\~ao Paulo, Brazil 05314-970}
\affiliation{University of Science and Technology of China, Hefei, Anhui 230026}
\affiliation{Shandong University, Qingdao, Shandong 266237}
\affiliation{Shanghai Institute of Applied Physics, Chinese Academy of Sciences, Shanghai 201800}
\affiliation{Southern Connecticut State University, New Haven, Connecticut 06515}
\affiliation{State University of New York, Stony Brook, New York 11794}
\affiliation{Instituto de Alta Investigaci\'on, Universidad de Tarapac\'a, Arica 1000000, Chile}
\affiliation{Temple University, Philadelphia, Pennsylvania 19122}
\affiliation{Texas A\&M University, College Station, Texas 77843}
\affiliation{University of Texas, Austin, Texas 78712}
\affiliation{Tsinghua University, Beijing 100084}
\affiliation{University of Tsukuba, Tsukuba, Ibaraki 305-8571, Japan}
\affiliation{Valparaiso University, Valparaiso, Indiana 46383}
\affiliation{Variable Energy Cyclotron Centre, Kolkata 700064, India}
\affiliation{Warsaw University of Technology, Warsaw 00-661, Poland}
\affiliation{Wayne State University, Detroit, Michigan 48201}
\affiliation{Yale University, New Haven, Connecticut 06520}

\author{M.~S.~Abdallah}\affiliation{American University of Cairo, New Cairo 11835, New Cairo, Egypt}
\author{J.~Adam}\affiliation{Brookhaven National Laboratory, Upton, New York 11973}
\author{L.~Adamczyk}\affiliation{AGH University of Science and Technology, FPACS, Cracow 30-059, Poland}
\author{J.~R.~Adams}\affiliation{Ohio State University, Columbus, Ohio 43210}
\author{J.~K.~Adkins}\affiliation{University of Kentucky, Lexington, Kentucky 40506-0055}
\author{G.~Agakishiev}\affiliation{Joint Institute for Nuclear Research, Dubna 141 980, Russia}
\author{I.~Aggarwal}\affiliation{Panjab University, Chandigarh 160014, India}
\author{M.~M.~Aggarwal}\affiliation{Panjab University, Chandigarh 160014, India}
\author{Z.~Ahammed}\affiliation{Variable Energy Cyclotron Centre, Kolkata 700064, India}
\author{I.~Alekseev}\affiliation{Alikhanov Institute for Theoretical and Experimental Physics NRC "Kurchatov Institute", Moscow 117218, Russia}\affiliation{National Research Nuclear University MEPhI, Moscow 115409, Russia}
% ...

\collaboration{STAR Collaboration}

\begin{abstract}
According to first principle Lattice QCD calculations, 
the transition from quark-gluon plasma to hadronic matter is a smooth crossover 
in the region $\mu_{\rm B}\leq T_{c}$. 
In this range the ratio, $C_{6}/C_{2}$, of net-baryon distributions are predicted to be negative. 
In this paper, we report the first measurement of the midrapidity net-proton $C_{6}/C_{2}$ 
from 27, 54.4 and 200~GeV Au+Au collisions at RHIC. 
The dependence on collision centrality and kinematic acceptance in ($p_{T}$, $y$) are analyzed. 
While for 27 and 54.4~GeV collisions the $C_{6}/C_{2}$ values are close to zero within uncertainties, 
it is observed that for 200~GeV collisions, the $C_{6}/C_{2}$ ratio becomes progressively 
negative from peripheral to central collisions.
Transport model calculations without critical dynamics predict 
mostly positive values except for the most central collisions within uncertainties. 
These observations seem to favor a smooth crossover in the high energy nuclear collisions at top RHIC energy.
\end{abstract}

\maketitle

\newcommand{\ave}[1]{\ensuremath{\langle#1\rangle} }
One of the main goals of high-energy nuclear physics is to understand 
the phase diagram of the strongly interacting matter predicted by quantum chromo-dynamics (QCD), 
with respect to temperature ($T$) and baryon chemical potential ($\mu_{\rm B}$). 
At high $T$ and/or $\mu_{\rm B}$, the strongly interacting matter called 
quark-gluon plasma (QGP) is predicted to exist, while the hadronic matter 
appears at low $T$~\cite{Fukushima:2010bq,BraunMunzinger:2009zz,Asakawa:1989bq}. 
The phase transition between QGP and hadronic matter 
at $\mu_{\rm B}\approx0$ was shown to be a smooth crossover, 
based on first-principle lattice QCD calculations~\cite{Aoki:2006we}.
At finite $\mu_{\rm B}$, on the other hand, the phase transition 
is predicted to be of the first order~\cite{Ejiri:2008xt}. 
If this is true, a critical end point may also exist, 
which is the connecting point between crossover and the 
first-order phase transition~\cite{Bowman:2008kc}.
\par
Experimentally, the QCD phase diagram can be explored by measuring heavy-ion collisions at various collision energies. 
The beam energy scan (BES) program was carried out at the Relativistic Heavy-Ion 
Collider (RHIC), and data sets for Au+Au collisions at $\sqrt{s_{\rm NN}}=$~7.7, 
11.5, 14.5, 19.6, 27, 39, 54.4, 62.4, and 200~GeV were collected by the STAR experiment.
The $r$th-order cumulants ($C_{r}$) and their ratios up to the 
fourth-order of net-charge, net-proton, and net-kaon multiplicity distributions 
have been measured to search for the 
critical point~\cite{Aggarwal:2010wy,net_proton,Adam:2020unf,net_charge,Adamczyk:2017wsl,STAR:2021iop}.
In particular, the ratio $C_{4}/C_{2}$ of the net-proton multiplicity distributions 
with an extended $p_{T}$ coverage shows a non-monotonic beam energy dependence 
in Au+Au central collisions~\cite{Adam:2020unf}. 
This is qualitatively consistent with a theoretical model 
prediction which incorporates a critical point~\cite{nonmonotonic,Stephanov:2011pb}.
Since the results are dominated by large statistical uncertainties at low 
collision energies, the beam energy scan phase II (BES-II) 
and the fixed-target programs are being performed 
to detect more definitive signatures of the critical point~\cite{BESII}. 
New findings on the QCD phase structure at large $\mu_{B}$ are thus 
expected in the near future from the BES-II program.

There is no direct experimental 
evidence of a smooth crossover at $\mu_{\rm B}\approx0$~MeV as predicted by the lattice 
QCD calculations. 
This can be studied by measuring the ratio of the sixth to 
second-order cumulant ($C_{6}/C_{2}$) of net-baryon and 
net-charge multiplicity distributions. 
According to the QCD model calculations,  
the $C_{6}/C_{2}$ values of net-baryon distributions become negative at $\sqrt{s_{\rm NN}}\geq$~60~GeV 
if the freeze-out, where ratios of particle yields are fixed,
occurs near the chiral crossover temperature~\cite{Friman}, whereas the hadron resonance gas model 
calculations yield a positive sign for $C_{6}/C_{2}$~\cite{Borsanyi:2018grb}. 
Recent model studies predict a negative sign of $C_{6}/C_{2}$ at $\sqrt{s_{\rm NN}}\geq7.7$~GeV 
within large uncertainties~\cite{Fu:2021oaw}. 
Furthermore, recent lattice QCD calculations also show a negative 
sign of the ratio of the sixth-order to the second-order baryon number 
susceptibilities, $\chi_{6}^{\rm B}/\chi_{2}^{\rm B}$, at the 
transition temperature for $\sqrt{s_{\rm NN}}\geq$~39~GeV~\cite{Borsanyi:2018grb,Bazavov:2020bjn}.
The susceptibility ratio can be compared to a corresponding ratio of experimentally measured cumulants. 

\par
This Letter reports $C_{6}/C_{2}$ of event-by-event net-proton 
multiplicity ($N_{p}-N_{\bar{p}}=\Delta N_{p}$) distributions 
for Au+Au collisions at $\sqrt{s_{\rm NN}}=$~27, 54.4, and 200~GeV.
These three collision energies correspond to $\mu_{\rm B}=$~144, 83, and 28~MeV, respectively, 
for the most central collisions~\cite{Adamczyk:2017iwn}.
The data sets for $\sqrt{s_{\rm NN}}=$~54.4 and 27~GeV were taken in 2017 and 2018. 
The data for 200~GeV were collected in 2010 and 2011. 
The numbers of events analyzed for 0-80\% centrality at $\sqrt{s_{\rm NN}}=$~27, 54.4, and 200~GeV 
are around $280$, $520$, and $900$ million, respectively. 
All data were taken with a minimum bias trigger, except for 
0-10\% centrality of 200~GeV data in 2010, which was taken with a special trigger 
with enhanced event samples for central collisions.
All data were taken with the time projection chamber (TPC) and the time of flight (TOF) 
detector at the solenoid tracker at RHIC (STAR). 
Events occurring within $|\Delta Z|<30$~cm from the center 
of the TPC along the beam line ($Z$-direction) are selected. 
The transverse positions of the collisions are required 
to be within $|\Delta R|<2$~cm to reject interactions between 
the beam and the beam pipe~\cite{Adamczyk:2017iwn}. 
Events from pileup, which is defined as the superposition of 
several single-collision events occurring within a small space and time interval,  
are rejected by cutting on the correlation between 
the charged particle multiplicity measured by the TPC and extrapolated 
tracks from the TPC to the hit positions in the TOF.

The collision centrality is defined using the charged particle multiplicities 
measured by the TPC in $|\eta|<1.0$.
Protons and antiprotons are excluded from the above centrality definition 
in order to minimize self-correlation effects~\cite{Luo:2017faz}. 
Event-by-event numbers of protons and antiprotons are measured 
at midrapidity, $|y|<0.5$, for the transverse momentum range $0.4<p_{T}\;({\rm GeV/c})<2.0$.  
Protons and antiprotons at $0.4<p_{T}\;({\rm GeV/c})<0.8$ are identified using ionization energy 
loss distributions measured by the TPC ($dE/dx$), 
while at $0.8<p_{T}\;({\rm GeV/c})<2.0$ 
they are identified using the mass squared ($m^{2}$) distributions measured by the TOF as well . 
The lower limit of the $p_{T}$ range is chosen to reject the secondary protons 
produced by interactions with the beam pipe.
Requiring the distance of closest approach to be less than 
$1$~cm with respect to the collision vertex suppresses the effect from 
the contribution of weak decay daughter protons. 
Weak decay protons which passed this cut are all included in the analysis.
Up to the fourth order, the effect of the decay is found to be small~\cite{Garg:2013ata}.
The purity of protons and antiprotons in the analyzed acceptance is 
higher than $95$\% for $\sqrt{s_{\rm NN}}=$~27, 54.4, and 200~GeV.

\par
\begin{figure}[htbp]
\begin{center}
\includegraphics[width=88mm]{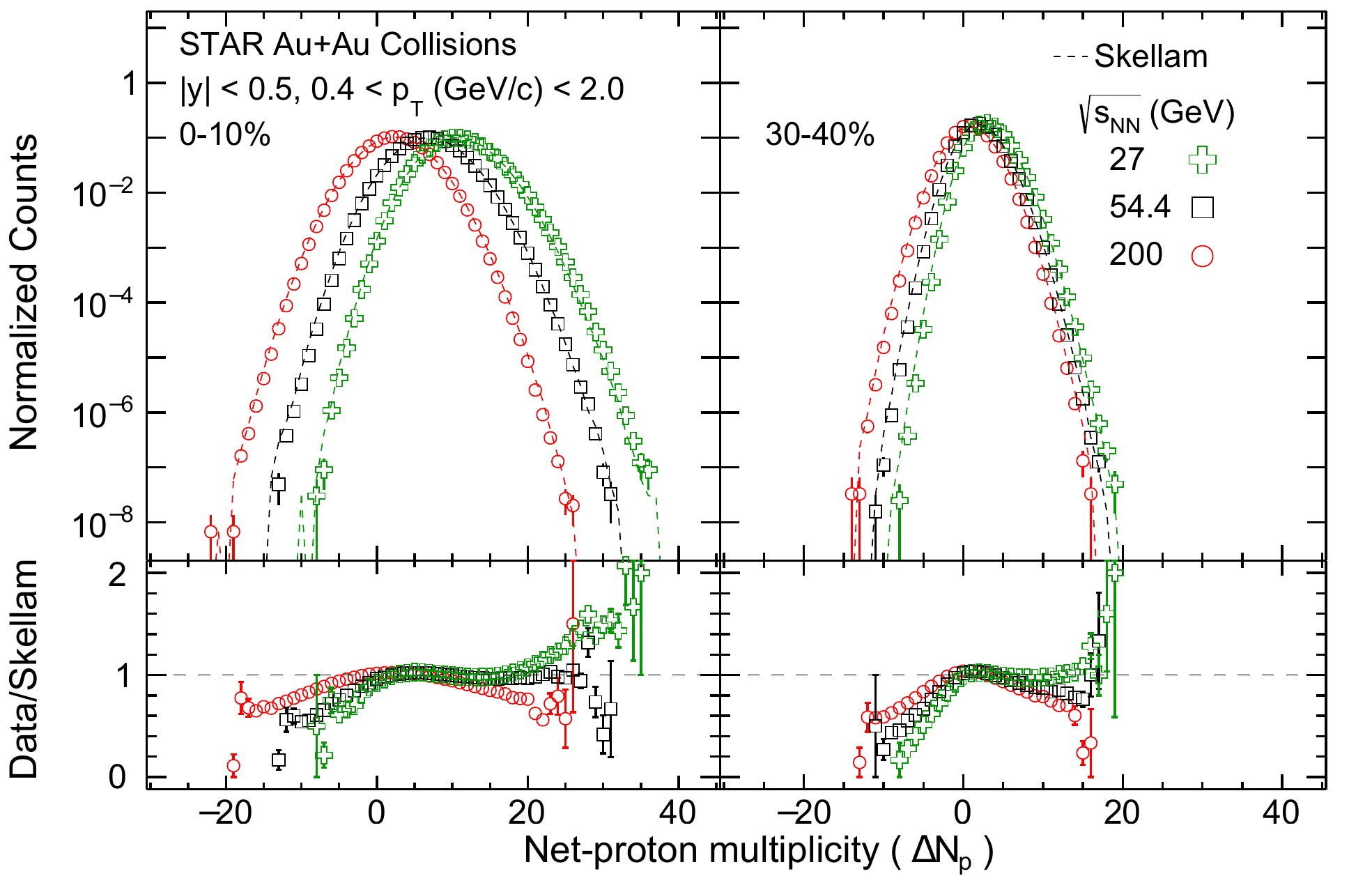}
\end{center}
\caption{
Event by event net-proton multiplicity, $\Delta{N_{\rm p}}$, distributions for 
Au+Au collisions at $\sqrt{s_{\rm NN}}=$~27, 54.4, and 200~GeV in 0-10\%(left) and 30-40\%(right) 
centralities at midrapidity ($|y|<0.5$) for the transverse 
momentum range of $0.4<p_{T}\;({\rm GeV/c})<2.0$. These distributions are normalized 
by the corresponding numbers of events and are not corrected for detector efficiencies. 
Statistical uncertainties are shown as vertical lines.
The dashed lines show the Skellam distributions for each collision energy and centrality.
The bottom panels show the ratio of the data to the Skellam expectations. 
}
\label{fig:Fig1}
\end{figure}

Event-by-event net-proton number ($\Delta N_{p}$) distributions for Au+Au collisions 
at $\sqrt{s_{\rm NN}}=$~27, 54.4, and 200~GeV for 0-10\% and 30-40\% centralities 
are shown in Fig.~\ref{fig:Fig1}. 
The plotted distributions are normalized by the corresponding 
total number of events and are not corrected for detector efficiencies.
The distributions for 0-10\% are broader than for 30-40\% as more 
protons and antiprotons are produced in central collisions. 
The shape of the distribution is characterized by various orders of cumulants~\cite{Asakawa:2015ybt}.
Definitions and formulas for cumulant calculations can be found in the Supplemental 
Material.
Cumulants are extensive variables proportional to 
the volume of the collision system~\cite{Asakawa:2015ybt}. 
This undesired volume effect is canceled by taking the ratio of 
different order cumulants.
Then the $C_{6}/C_{2}$ value can be compared with the ratio of baryon number susceptibilities 
($\chi^{\rm B}_{6}/\chi^{\rm B}_{2}$) from lattice QCD calculations, 
keeping in mind the difference between net-baryon from theory calculations and net-proton from the data within 
the limited experimental acceptance.
When multiplicities of protons and antiprotons follow Poisson distributions, 
the resulting net-proton distribution is called a Skellam distribution. 
The odd-order and even-order cumulants of the Skellam distribution are expressed 
by the difference and sum of the mean values ($C_{1}$) of the Poisson distributions, respectively.
Hence, $C_{6}/C_{2}=1$ for the Skellam distribution. 
The Skellam distributions determined from $C_{1}$ of protons and antiprotons 
for each collision energy and centrality are shown by dashed lines in Fig.~\ref{fig:Fig1}.
According to the ratio of data to the Skellam expectations, shown in the lower part of Fig.~\ref{fig:Fig1}, deviations from 
the Skellam distributions are seen especially at the tails of the distribution. 

It is known that the statistical uncertainties 
on higher-order cumulants become larger for broader 
distributions~\cite{Luo:2017faz}. 
A model study indicates that higher-order cumulants
suffer from larger statistical uncertainties.  
The effect increases with increasing order of the cumulant~\cite{eff_xiaofeng}.
Statistical uncertainties also depend on the detector efficiencies.
A lower efficiency gives larger statistical errors for cumulants 
after efficiency corrections.

A centrality bin width correction is applied 
for each centrality bin to suppress the effect from the initial volume 
fluctuations~\cite{Skokov:2012ds,Luo:2017faz,Sugiura:2019toh}.
Statistical uncertainties are calculated 
using a bootstrap method~\cite{Luo:2017faz,Pandav:2018bdx}.

All results of $C_{6}/C_{2}$ presented in this Letter are corrected for the detector efficiency 
assuming that the detector efficiencies 
follow the binomial distribution~\cite{eff_kitazawa,eff_koch,tsukuba_eff_separate,eff_psd_volker,eff_xiaofeng,Luo:2018ofd,eff_psd_kitazawa,Nonaka:2017kko}.
Non-binomial efficiencies~\cite{binomial_breaking} are also studied 
through detector simulations in the STAR environment.  
Cumulants are corrected for non-binomial efficiencies using the unfolding 
and moment expansion approaches~\cite{Esumi:2020xdo,Nonaka:2018mgw}.
Results up to the sixth-order cumulant for Au+Au central collisions 
at $\sqrt{s_{\rm NN}}=200$~GeV are presented in the Supplemental Material. 
It is concluded that the results corrected for non-binomial efficiencies 
are consistent with the results from the binomial efficiency correction within 
statistical uncertainties. 

Systematic uncertainties are estimated by changing the following 
variables used to select protons and antiprotons: 
the distance of closest approach to the primary collision vertex and  
number of hits in the TPC to reconstruct tracks for the track quality cut, 
$dE/dx$, and $m^{2}$ selections for (anti)proton identification criteria. 
A Barlow check is done to remove the statistical effects from 
being counted as part of systematic uncertainties~\cite{Barlow:2002yb}.
The contribution from track quality cuts is dominant for central collisions. 
The systematic uncertainties from each source go down below 10\% in peripheral collisions.
The uncertainties for each source are then added in quadrature. 
The total systematic uncertainties are  87\%, 70\%, and 37\% at 27, 54.4, and 200~GeV, respectively, 
for 0-10\% central collisions, and the corresponding totals decrease down to a few percent in peripheral collisions.

\begin{figure}[htbp]
\begin{center}
\includegraphics[width=87mm]{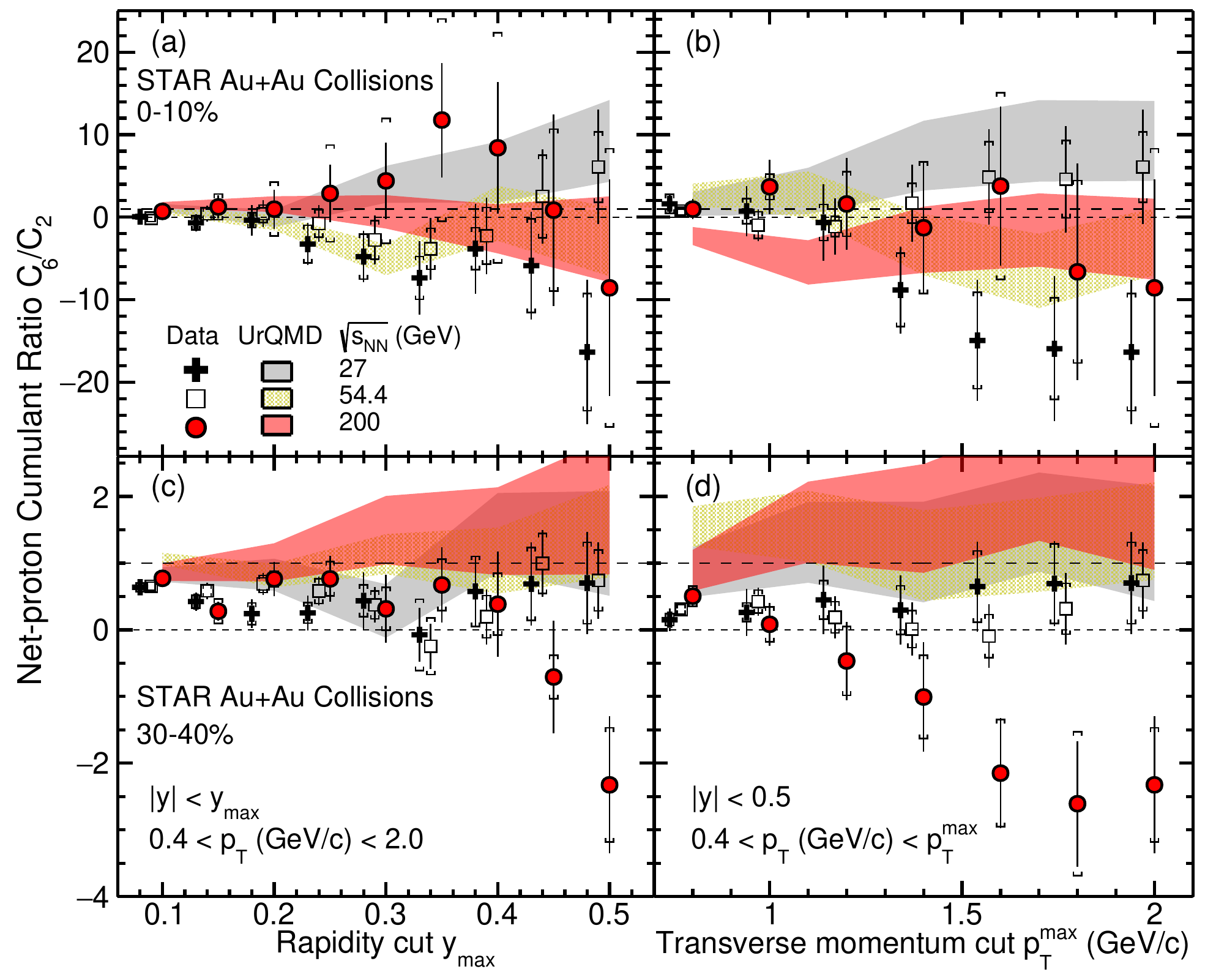}
\end{center}
\caption{
Net-proton $C_{6}/C_{2}$ as a function of rapidity (left) 
and transverse momentum acceptance (right) from $\sqrt{s_{\rm NN}}=27$ (crosses), 
54.4 (open squares), and 200~GeV (filled circles) Au+Au collisions. 
The upper and lower plots are for 0-10\% and 30-40\% centralities, respectively. 
The error bars are statistical and caps are systematic errors.
Points for different beam energies are staggered horizontally to improve clarity.
UrQMD transport model results are shown as shaded and hatched bands. 
The Skellam expectation ($C_{6}/C_{2}=1$) is shown as long-dashed lines. 
 }
\label{fig:Fig2}
\end{figure}

Figure~\ref{fig:Fig2} shows the net-proton $C_{6}/C_{2}$ for Au+Au collisions 
for 0-10\% and 30-40\% centralities at $\sqrt{s_{\rm NN}}=$~27, 54.4, and 200~GeV 
as a function of rapidity and $p_{T}$ acceptance.
The values of $C_{6}/C_{2}$ approach the Skellam expectation, 
$C_{6}/C_{2}=1$, with narrow acceptance in $p_{T}$ and rapidity. 
The reason is that multiplicity distributions of 
protons and antiprotons are close to the Poisson distribution 
because of lower particle multiplicity and thus less correlations, 
and therefore the observed $C_{6}/C_{2}$ is dominated by statistical fluctuations.
The fraction of measured protons to total protons 
integrated in whole $p_{T}$ region is 33\% 
for $0.4<p_{T}\;({\rm GeV/c})<0.8$ and 86\% for $0.4<p_{T}\;({\rm GeV/c})<2.0$ 
at 200~GeV. 
Although the $C_{6}/C_{2}$ values at the smallest acceptance 
of $|y|<0.1$ or $0.4<p_{T}\;({\rm GeV/c})<0.8$ 
in Fig.~\ref{fig:Fig2} are still smaller than unity, we have checked that 
the results are consistent with unity with further narrowed acceptance.
The $C_{6}/C_{2}$ values for 0-10\% centrality decrease as the acceptance is increased 
at 27~GeV, while $C_{6}/C_{2}$ is 
nearly constant for 54.4 and 200~GeV within uncertainties.
On the other hand, the results for 30-40\% centrality show a strong decrease with increasing acceptance at 200~GeV and 
are almost flat for 27 and 54.4~GeV.
Results from the transport model UrQMD~\cite{Bleicher:1999xi}, in which hadronic 
interactions are dominant and there is no phase transition implemented, are shown 
by shaded and hatched-bands in Fig.~\ref{fig:Fig2}. 
The event statistics used in the UrQMD calculations are 215, 100, and 95 million for 27, 54.4, and 
200~GeV minimum bias Au+Au collisions, respectively. All experimental cuts 
in terms of the collision centrality, 
rapidity, and transverse momentum acceptance are applied in the calculations.  
The $C_{6}/C_{2}$ values from UrQMD are flat as a function of rapidity and $p_{T}$ acceptance at 27 and 200~GeV, while 
the sign changes for central collisions at 54.4~GeV albeit with large uncertainties.
Note that the thermal blurring in rapidity for conserved charges is discussed in Ref.~\cite{Ohnishi:2016bdf}. 
More studies are necessary in order to understand the rapidity dependence as a function of collision energies. 

\begin{figure}[htbp]
\begin{center}
\includegraphics[width=85mm]{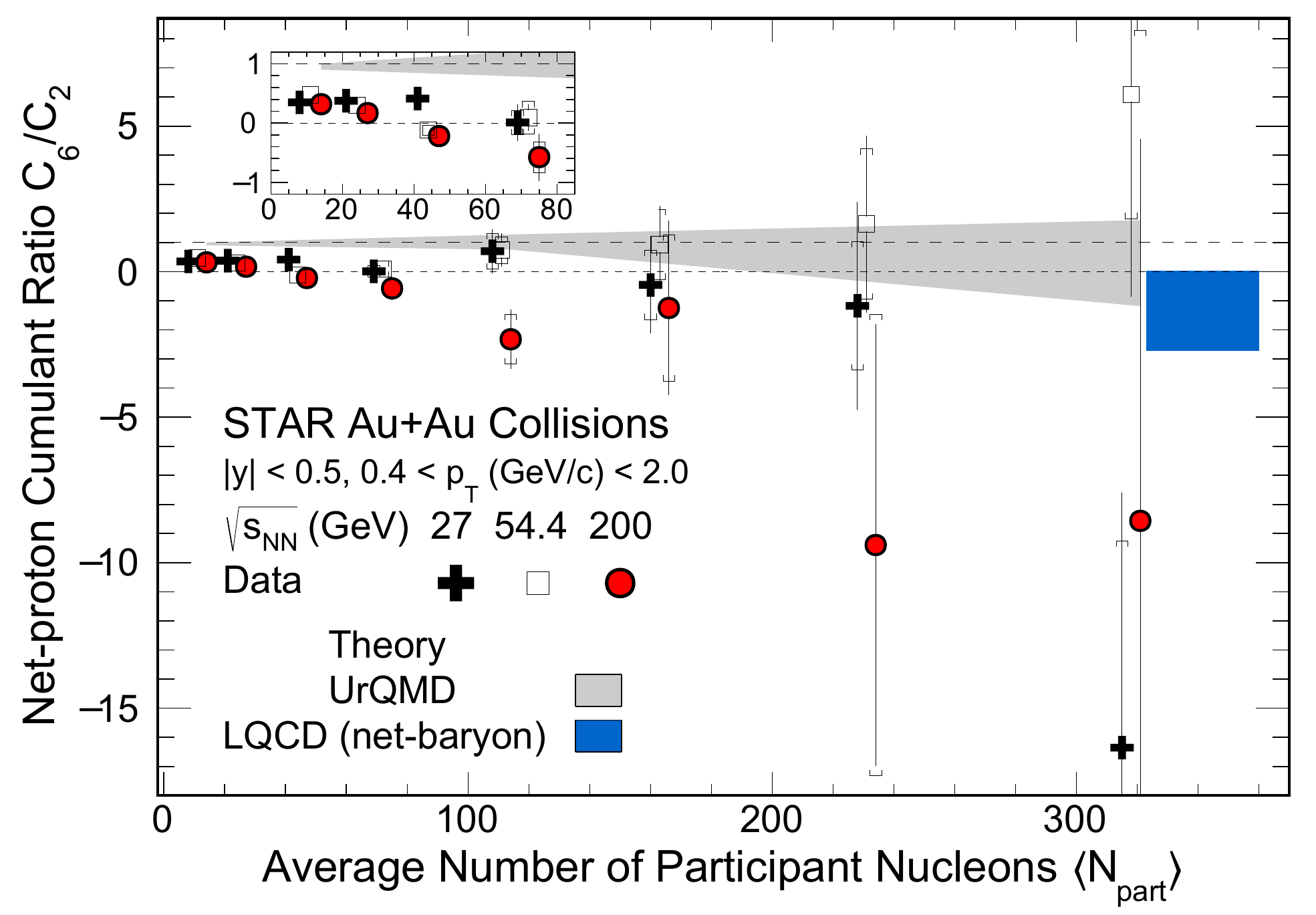}
\end{center}
\caption{ 
Collision centrality dependence of net-proton 
$C_{6}/C_{2}$ in Au+Au collisions for $\sqrt{s_{\rm NN}}=$~27, 54.4, and 200~GeV 
within $|y|<0.5$ and $0.4<p_{T}\;({\rm GeV/c})<2.0$. 
The error bars are statistical and caps are systematic errors. 
Points for different beam energies are staggered horizontally to improve clarity.
A shaded band shows the results from UrQMD model calculations. 
UrQMD calculations from the above three collision energies are consistent among them so they are merged in order to reduce statistical fluctuations.
Details on these calculations can be found in the Supplemental 
Material.
The lattice QCD calculations on net-baryon number fluctuations~\cite{Borsanyi:2018grb,Bazavov:2020bjn} 
for $T=160$~MeV and $\mu_{\rm B}<$~110~MeV. 
are shown as a blue band at $\ave{N_{\rm part}}\approx340$.
The inset shows the expanded region of 40-80\% centrality.
}
\label{fig:Fig3}
\end{figure}

In Fig.~\ref{fig:Fig3}, the centrality dependence of the net-proton $C_{6}/C_{2}$ at midrapidity
is shown for all three collision energies. The data with the largest number of 
participant nucleons ($N_{\rm part}$) corresponds to the top 0-10\% central collisions 
and the rest of the points are for 10-20\%, 20-30\%, 30-40\%,... , and 70-80\% centralities. 
For 200~GeV collisions (filled circles), $C_{6}/C_{2}$ values are approaching the Skellam expectation ($C_{6}/C_{2}=1$) 
from central to peripheral collisions. 
The $C_{6}/C_{2}$ values then start to be negative from 50-60\% centrality, and stay negative systematically  
in central collisions within large uncertainties. 
Most $C_{6}/C_{2}$ measurements at 27 and 54.4~GeV are consistent with zero within uncertainties. 
We find that the $C_{6}/C_{2}$ values are negative within $1.7$ sigma at 200~GeV 30-40\% centrality 
with statistical and systematic uncertainties added in quadrature.

QCD-inspired model calculations~\cite{Friman} show that at vanishing baryon chemical potential, 
the crossover transition from the QGP to hadronic phase and its 
remnants will affect higher-order cumulant ratios. 
The model further suggests that the value of $C_{6}/C_{2}$ of the net-baryon distribution should be positive and negative 
for the emerging medium from hadronic and QGP phases, respectively. 
All of the results from the UrQMD calculations are consistent 
with the Skellam expectation ($C_{6}/C_{2}=1$) within large statistical fluctuations towards more central collisions.  
Overall, the UrQMD calculations of the net-proton $C_{6}/C_{2}$ reproduce, within errors, 
the measured centrality dependence for 27 and 54.4~GeV Au+Au collisions. 
Experimental results for 200~GeV are below UrQMD calculations systematically from peripheral to central collisions.  

First-principle lattice QCD calculations offer accurate information on the properties 
of a thermalized system with zero baryon chemical potential. For example, see Ref.~\cite{Aoki:2006we}. 
By a Taylor expansion at small $\mu_{B}$, one could extend the predictions to finite 
values of the baryon chemical potential. The lattice calculations with a temperature 
of 160~MeV and baryon chemical potential up to $\mu_{B}\sim110$~MeV  
are shown as the blue band in Fig.~\ref{fig:Fig3}~\cite{Bazavov:2020bjn,Borsanyi:2018grb}. 
Lattice calculations also indicate that in the region of $\mu_{B}/T< 1$, 
the transition from QGP to hadronic matter is a smooth crossover~\cite{Borsanyi:2018grb}. 
The $\mu_{\rm B}/T$ ratios are approximately 0.17,  0.55, and 0.93 
for central Au+Au collisions at $\sqrt{s_{\rm NN}}=200$, 54, and 27~GeV, respectively.  
Please note there are caveats 
when comparing experimental data with lattice calculations. While the current experimental data are 
midrapidity net-proton distributions from the kinematic region of $|y|<0.5$ and $0.4<p_{T}\;({\rm GeV/c})<2.0$, 
the lattice results are for net-baryons and do not incorporate any of the experimental kinematic cuts~\cite{eff_kitazawa}. 
It is also known that the cumulants are affected by both baryon number conservation and baryon stopping~\cite{Bzdak:2012an,Braun-Munzinger:2016yjz,Vovchenko:2020kwg,Vovchenko:2021kxx}
which are expected to be more significant towards lower collision energies~\cite{Braun-Munzinger:2020jbk,Mishra:2017mor}. 
Both effects are present in the results presented here.
In addition, the lattice simulates a thermalized system but without other dynamics such as 
collective expansion in high-energy nuclear collisions. 
These differences between experiments and lattice calculations 
need to be carefully handled in the future for a quantitative comparison. 
With these caveats in mind, the trend observed in Au+Au collisions at 200~GeV 
that $C_{6}/C_{2}$ becomes negative with increasing centrality is potentially consistent 
with the smooth crossover scenario. 
Higher statistics data sets are necessary in order to establish trend in the finite $\mu_{\rm B}$ range.

\par
In summary, we report the first measurements of the net-proton higher-order 
cumulant ratio $C_{6}/C_{2}$ from 27, 54.4 and 200~GeV Au+Au collisions measured 
by the STAR detector at RHIC. The data is taken from the kinematic region ($|y|<0.5$ and $0.4<p_{T}\;({\rm GeV/c})<2.0$). 
Data from 200~GeV collisions are found to be negative progressively 
in more central collisions within the maximum acceptance, while the ratios 
from 27 and 54.4~GeV collisions are found to be close to zero within uncertainties. 
Without critical dynamics, the transport model UrQMD calculations predict the ratio $C_{6}/C_{2}$ 
around a statistical baseline in all cases. 
Lattice QCD calculations, with $T=160$~MeV and $\mu_{\rm B}=0$-$110$~MeV, 
predict the negative value of $C_{6}/C_{2}\sim-1.5$, which is qualitatively consistent with the experimental 
results of central Au+Au collisions at top RHIC energies. 
These new measurements are statistics limited and seem to favor a smooth crossover for the QGP-hadronic matter transition. 
Future measurements with high statistics will provide more detailed information 
about the phase structure at the low baryon density region.

\par
We thank the RHIC Operations Group and RCF at BNL, the NERSC Center at LBNL, and the Open Science Grid consortium for providing resources and support.  This work was supported in part by the Office of Nuclear Physics within the U.S. DOE Office of Science, the U.S. National Science Foundation, the Ministry of Education and Science of the Russian Federation, National Natural Science Foundation of China, Chinese Academy of Science, the Ministry of Science and Technology of China and the Chinese Ministry of Education, the Higher Education Sprout Project by Ministry of Education at NCKU, the National Research Foundation of Korea, Czech Science Foundation and Ministry of Education, Youth and Sports of the Czech Republic, Hungarian National Research, Development and Innovation Office, New National Excellency Programme of the Hungarian Ministry of Human Capacities, Department of Atomic Energy and Department of Science and Technology of the Government of India, the National Science Centre of Poland, the Ministry  of Science, Education and Sports of the Republic of Croatia, RosAtom of Russia and German Bundesministerium fur Bildung, Wissenschaft, Forschung and Technologie (BMBF), Helmholtz Association, Ministry of Education, Culture, Sports, Science, and Technology (MEXT) and Japan Society for the Promotion of Science (JSPS).

\bibliography{ForArxiv}%

\providecommand{\noopsort}[1]{}\providecommand{\singleletter}[1]{#1}%
\begin{thebibliography}{46}%
\makeatletter
\providecommand \@ifxundefined [1]{%
 \@ifx{#1\undefined}
}%
\providecommand \@ifnum [1]{%
 \ifnum #1\expandafter \@firstoftwo
 \else \expandafter \@secondoftwo
 \fi
}%
\providecommand \@ifx [1]{%
 \ifx #1\expandafter \@firstoftwo
 \else \expandafter \@secondoftwo
 \fi
}%
\providecommand \natexlab [1]{#1}%
\providecommand \enquote  [1]{``#1''}%
\providecommand \bibnamefont  [1]{#1}%
\providecommand \bibfnamefont [1]{#1}%
\providecommand \citenamefont [1]{#1}%
\providecommand \href@noop [0]{\@secondoftwo}%
\providecommand \href [0]{\begingroup \@sanitize@url \@href}%
\providecommand \@href[1]{\@@startlink{#1}\@@href}%
\providecommand \@@href[1]{\endgroup#1\@@endlink}%
\providecommand \@sanitize@url [0]{\catcode `\\12\catcode `\$12\catcode
  `\&12\catcode `\#12\catcode `\^12\catcode `\_12\catcode `\%12\relax}%
\providecommand \@@startlink[1]{}%
\providecommand \@@endlink[0]{}%
\providecommand \url  [0]{\begingroup\@sanitize@url \@url }%
\providecommand \@url [1]{\endgroup\@href {#1}{\urlprefix }}%
\providecommand \urlprefix  [0]{URL }%
\providecommand \Eprint [0]{\href }%
\providecommand \doibase [0]{http://dx.doi.org/}%
\providecommand \selectlanguage [0]{\@gobble}%
\providecommand \bibinfo  [0]{\@secondoftwo}%
\providecommand \bibfield  [0]{\@secondoftwo}%
\providecommand \translation [1]{[#1]}%
\providecommand \BibitemOpen [0]{}%
\providecommand \bibitemStop [0]{}%
\providecommand \bibitemNoStop [0]{.\EOS\space}%
\providecommand \EOS [0]{\spacefactor3000\relax}%
\providecommand \BibitemShut  [1]{\csname bibitem#1\endcsname}%
\let\auto@bib@innerbib\@empty
\bibitem [{\citenamefont {Fukushima}\ and\ \citenamefont
  {Hatsuda}(2011)}]{Fukushima:2010bq}%
  \BibitemOpen
  \bibfield  {author} {\bibinfo {author} {\bibfnamefont {K.}~\bibnamefont
  {Fukushima}}\ and\ \bibinfo {author} {\bibfnamefont {T.}~\bibnamefont
  {Hatsuda}},\ }\href {\doibase 10.1088/0034-4885/74/1/014001} {\bibfield
  {journal} {\bibinfo  {journal} {Rept. Prog. Phys.}\ }\textbf {\bibinfo
  {volume} {74}},\ \bibinfo {pages} {014001} (\bibinfo {year}
  {2011})}\BibitemShut {NoStop}%
\bibitem [{\citenamefont {Braun-Munzinger}\ and\ \citenamefont
  {Wambach}(2009)}]{BraunMunzinger:2009zz}%
  \BibitemOpen
  \bibfield  {author} {\bibinfo {author} {\bibfnamefont {P.}~\bibnamefont
  {Braun-Munzinger}}\ and\ \bibinfo {author} {\bibfnamefont {J.}~\bibnamefont
  {Wambach}},\ }\href {\doibase 10.1103/RevModPhys.81.1031} {\bibfield
  {journal} {\bibinfo  {journal} {Rev. Mod. Phys.}\ }\textbf {\bibinfo {volume}
  {81}},\ \bibinfo {pages} {1031} (\bibinfo {year} {2009})}\BibitemShut
  {NoStop}%
\bibitem [{\citenamefont {Asakawa}\ and\ \citenamefont
  {Yazaki}(1989)}]{Asakawa:1989bq}%
  \BibitemOpen
  \bibfield  {author} {\bibinfo {author} {\bibfnamefont {M.}~\bibnamefont
  {Asakawa}}\ and\ \bibinfo {author} {\bibfnamefont {K.}~\bibnamefont
  {Yazaki}},\ }\href {\doibase 10.1016/0375-9474(89)90002-X} {\bibfield
  {journal} {\bibinfo  {journal} {Nucl. Phys. A}\ }\textbf {\bibinfo {volume}
  {504}},\ \bibinfo {pages} {668} (\bibinfo {year} {1989})}\BibitemShut
  {NoStop}%
\bibitem [{\citenamefont {Aoki}\ \emph {et~al.}(2006)\citenamefont {Aoki},
  \citenamefont {Endrodi}, \citenamefont {Fodor}, \citenamefont {Katz},\ and\
  \citenamefont {Szabo}}]{Aoki:2006we}%
  \BibitemOpen
  \bibfield  {author} {\bibinfo {author} {\bibfnamefont {Y.}~\bibnamefont
  {Aoki}}, \bibinfo {author} {\bibfnamefont {G.}~\bibnamefont {Endrodi}},
  \bibinfo {author} {\bibfnamefont {Z.}~\bibnamefont {Fodor}}, \bibinfo
  {author} {\bibfnamefont {S.~D.}\ \bibnamefont {Katz}}, \ and\ \bibinfo
  {author} {\bibfnamefont {K.~K.}\ \bibnamefont {Szabo}},\ }\href {\doibase
  10.1038/nature05120} {\bibfield  {journal} {\bibinfo  {journal} {Nature}\
  }\textbf {\bibinfo {volume} {443}},\ \bibinfo {pages} {675} (\bibinfo {year}
  {2006})}\BibitemShut {NoStop}%
\bibitem [{\citenamefont {Ejiri}(2008)}]{Ejiri:2008xt}%
  \BibitemOpen
  \bibfield  {author} {\bibinfo {author} {\bibfnamefont {S.}~\bibnamefont
  {Ejiri}},\ }\href {\doibase 10.1103/PhysRevD.78.074507} {\bibfield  {journal}
  {\bibinfo  {journal} {Phys. Rev.}\ }\textbf {\bibinfo {volume} {D78}},\
  \bibinfo {pages} {074507} (\bibinfo {year} {2008})}\BibitemShut {NoStop}%
\bibitem [{\citenamefont {Bowman}\ and\ \citenamefont
  {Kapusta}(2009)}]{Bowman:2008kc}%
  \BibitemOpen
  \bibfield  {author} {\bibinfo {author} {\bibfnamefont {E.~S.}\ \bibnamefont
  {Bowman}}\ and\ \bibinfo {author} {\bibfnamefont {J.~I.}\ \bibnamefont
  {Kapusta}},\ }\href {\doibase 10.1103/PhysRevC.79.015202} {\bibfield
  {journal} {\bibinfo  {journal} {Phys. Rev.}\ }\textbf {\bibinfo {volume}
  {C79}},\ \bibinfo {pages} {015202} (\bibinfo {year} {2009})}\BibitemShut
  {NoStop}%
\bibitem [{\citenamefont {Aggarwal}\ \emph {et~al.}(2010)\citenamefont
  {Aggarwal} \emph {et~al.}}]{Aggarwal:2010wy}%
  \BibitemOpen
  \bibfield  {author} {\bibinfo {author} {\bibfnamefont {M.~M.}\ \bibnamefont
  {Aggarwal}} \emph {et~al.} (\bibinfo {collaboration} {STAR collaboration}),\
  }\href {\doibase 10.1103/PhysRevLett.105.022302} {\bibfield  {journal}
  {\bibinfo  {journal} {Phys. Rev. Lett.}\ }\textbf {\bibinfo {volume} {105}},\
  \bibinfo {pages} {022302} (\bibinfo {year} {2010})}\BibitemShut {NoStop}%
\bibitem [{\citenamefont {Adamczyk}\ \emph
  {et~al.}(2014{\natexlab{a}})\citenamefont {Adamczyk} \emph
  {et~al.}}]{net_proton}%
  \BibitemOpen
  \bibfield  {author} {\bibinfo {author} {\bibfnamefont {L.}~\bibnamefont
  {Adamczyk}} \emph {et~al.} (\bibinfo {collaboration} {STAR collaboration}),\
  }\href {\doibase 10.1103/PhysRevLett.112.032302} {\bibfield  {journal}
  {\bibinfo  {journal} {Phys. Rev. Lett.}\ }\textbf {\bibinfo {volume} {112}},\
  \bibinfo {pages} {032302} (\bibinfo {year} {2014}{\natexlab{a}})}\BibitemShut
  {NoStop}%
\bibitem [{\citenamefont {Adam}\ \emph {et~al.}(2021)\citenamefont {Adam} \emph
  {et~al.}}]{Adam:2020unf}%
  \BibitemOpen
  \bibfield  {author} {\bibinfo {author} {\bibfnamefont {J.}~\bibnamefont
  {Adam}} \emph {et~al.} (\bibinfo {collaboration} {STAR collaboration}),\
  }\href {\doibase 10.1103/PhysRevLett.126.092301} {\bibfield  {journal}
  {\bibinfo  {journal} {Phys. Rev. Lett.}\ }\textbf {\bibinfo {volume} {126}},\
  \bibinfo {pages} {092301} (\bibinfo {year} {2021})}\BibitemShut {NoStop}%
\bibitem [{\citenamefont {Adamczyk}\ \emph
  {et~al.}(2014{\natexlab{b}})\citenamefont {Adamczyk} \emph
  {et~al.}}]{net_charge}%
  \BibitemOpen
  \bibfield  {author} {\bibinfo {author} {\bibfnamefont {L.}~\bibnamefont
  {Adamczyk}} \emph {et~al.} (\bibinfo {collaboration} {STAR collaboration}),\
  }\href {\doibase 10.1103/PhysRevLett.113.092301} {\bibfield  {journal}
  {\bibinfo  {journal} {Phys. Rev. Lett.}\ }\textbf {\bibinfo {volume} {113}},\
  \bibinfo {pages} {092301} (\bibinfo {year} {2014}{\natexlab{b}})}\BibitemShut
  {NoStop}%
\bibitem [{\citenamefont {Adamczyk}\ \emph {et~al.}(2018)\citenamefont
  {Adamczyk} \emph {et~al.}}]{Adamczyk:2017wsl}%
  \BibitemOpen
  \bibfield  {author} {\bibinfo {author} {\bibfnamefont {L.}~\bibnamefont
  {Adamczyk}} \emph {et~al.} (\bibinfo {collaboration} {STAR collaboration}),\
  }\href {\doibase 10.1016/j.physletb.2018.07.066} {\bibfield  {journal}
  {\bibinfo  {journal} {Phys. Lett. B}\ }\textbf {\bibinfo {volume} {785}},\
  \bibinfo {pages} {551} (\bibinfo {year} {2018})}\BibitemShut {NoStop}%
\bibitem [{\citenamefont {Abdallah}\ \emph {et~al.}(2021)\citenamefont
  {Abdallah} \emph {et~al.}}]{STAR:2021iop}%
  \BibitemOpen
  \bibfield  {author} {\bibinfo {author} {\bibfnamefont {M.}~\bibnamefont
  {Abdallah}} \emph {et~al.} (\bibinfo {collaboration} {STAR}),\ }\href
  {\doibase 10.1103/PhysRevC.104.024902} {\bibfield  {journal} {\bibinfo
  {journal} {Phys. Rev. C}\ }\textbf {\bibinfo {volume} {104}},\ \bibinfo
  {pages} {024902} (\bibinfo {year} {2021})}\BibitemShut {NoStop}%
\bibitem [{\citenamefont {Stephanov}\ \emph {et~al.}(1999)\citenamefont
  {Stephanov}, \citenamefont {Rajagopal},\ and\ \citenamefont
  {Shuryak}}]{nonmonotonic}%
  \BibitemOpen
  \bibfield  {author} {\bibinfo {author} {\bibfnamefont {M.~A.}\ \bibnamefont
  {Stephanov}}, \bibinfo {author} {\bibfnamefont {K.}~\bibnamefont
  {Rajagopal}}, \ and\ \bibinfo {author} {\bibfnamefont {E.~V.}\ \bibnamefont
  {Shuryak}},\ }\href {\doibase 10.1103/PhysRevD.60.114028} {\bibfield
  {journal} {\bibinfo  {journal} {Phys. Rev.}\ }\textbf {\bibinfo {volume}
  {D60}},\ \bibinfo {pages} {114028} (\bibinfo {year} {1999})}\BibitemShut
  {NoStop}%
\bibitem [{\citenamefont {Stephanov}(2011)}]{Stephanov:2011pb}%
  \BibitemOpen
  \bibfield  {author} {\bibinfo {author} {\bibfnamefont {M.}~\bibnamefont
  {Stephanov}},\ }\href {\doibase 10.1103/PhysRevLett.107.052301} {\bibfield
  {journal} {\bibinfo  {journal} {Phys. Rev. Lett.}\ }\textbf {\bibinfo
  {volume} {107}},\ \bibinfo {pages} {052301} (\bibinfo {year}
  {2011})}\BibitemShut {NoStop}%
\bibitem [{BES()}]{BESII}%
  \BibitemOpen
  \href@noop {} {}\bibinfo {howpublished}
  {\url{https://drupal.star.bnl.gov/STAR/starnotes/public/sn0598}},\ \bibinfo
  {note} {{\sc BES-II} white paper: STAR Note}\BibitemShut {NoStop}%
\bibitem [{\citenamefont {Friman}\ \emph {et~al.}(2011)\citenamefont {Friman},
  \citenamefont {Karsch}, \citenamefont {Redlich},\ and\ \citenamefont
  {Skokov}}]{Friman}%
  \BibitemOpen
  \bibfield  {author} {\bibinfo {author} {\bibfnamefont {B.}~\bibnamefont
  {Friman}}, \bibinfo {author} {\bibfnamefont {F.}~\bibnamefont {Karsch}},
  \bibinfo {author} {\bibfnamefont {K.}~\bibnamefont {Redlich}}, \ and\
  \bibinfo {author} {\bibfnamefont {V.}~\bibnamefont {Skokov}},\ }\href
  {\doibase 10.1140/epjc/s10052-011-1694-2} {\bibfield  {journal} {\bibinfo
  {journal} {Eur. Phys. J.}\ }\textbf {\bibinfo {volume} {C71}},\ \bibinfo
  {pages} {1694} (\bibinfo {year} {2011})}\BibitemShut {NoStop}%
\bibitem [{\citenamefont {Borsanyi}\ \emph {et~al.}(2018)\citenamefont
  {Borsanyi}, \citenamefont {Fodor}, \citenamefont {Guenther}, \citenamefont
  {Katz}, \citenamefont {Pasztor}, \citenamefont {Portillo}, \citenamefont
  {Ratti},\ and\ \citenamefont {Szabo}}]{Borsanyi:2018grb}%
  \BibitemOpen
  \bibfield  {author} {\bibinfo {author} {\bibfnamefont {S.}~\bibnamefont
  {Borsanyi}}, \bibinfo {author} {\bibfnamefont {Z.}~\bibnamefont {Fodor}},
  \bibinfo {author} {\bibfnamefont {J.~N.}\ \bibnamefont {Guenther}}, \bibinfo
  {author} {\bibfnamefont {S.~K.}\ \bibnamefont {Katz}}, \bibinfo {author}
  {\bibfnamefont {A.}~\bibnamefont {Pasztor}}, \bibinfo {author} {\bibfnamefont
  {I.}~\bibnamefont {Portillo}}, \bibinfo {author} {\bibfnamefont
  {C.}~\bibnamefont {Ratti}}, \ and\ \bibinfo {author} {\bibfnamefont {K.~K.}\
  \bibnamefont {Szabo}},\ }\href@noop {} {\bibfield  {journal} {\bibinfo
  {journal} {JHEP}\ }\textbf {\bibinfo {volume} {10}},\ \bibinfo {pages} {205}
  (\bibinfo {year} {2018})}\BibitemShut {NoStop}%
\bibitem [{\citenamefont {Fu}\ \emph {et~al.}(2021)\citenamefont {Fu},
  \citenamefont {Luo}, \citenamefont {Pawlowski}, \citenamefont {Rennecke},
  \citenamefont {Wen},\ and\ \citenamefont {Yin}}]{Fu:2021oaw}%
  \BibitemOpen
  \bibfield  {author} {\bibinfo {author} {\bibfnamefont {W.-j.}\ \bibnamefont
  {Fu}}, \bibinfo {author} {\bibfnamefont {X.}~\bibnamefont {Luo}}, \bibinfo
  {author} {\bibfnamefont {J.~M.}\ \bibnamefont {Pawlowski}}, \bibinfo {author}
  {\bibfnamefont {F.}~\bibnamefont {Rennecke}}, \bibinfo {author}
  {\bibfnamefont {R.}~\bibnamefont {Wen}}, \ and\ \bibinfo {author}
  {\bibfnamefont {S.}~\bibnamefont {Yin}},\ }\href {\doibase
  10.1103/PhysRevD.104.094047} {\bibfield  {journal} {\bibinfo  {journal}
  {Phys. Rev. D}\ }\textbf {\bibinfo {volume} {104}},\ \bibinfo {pages}
  {094047} (\bibinfo {year} {2021})}\BibitemShut {NoStop}%
\bibitem [{\citenamefont {Bazavov}\ \emph {et~al.}(2020)\citenamefont {Bazavov}
  \emph {et~al.}}]{Bazavov:2020bjn}%
  \BibitemOpen
  \bibfield  {author} {\bibinfo {author} {\bibfnamefont {A.}~\bibnamefont
  {Bazavov}} \emph {et~al.},\ }\href {\doibase 10.1103/PhysRevD.101.074502}
  {\bibfield  {journal} {\bibinfo  {journal} {Phys. Rev.}\ }\textbf {\bibinfo
  {volume} {D101}},\ \bibinfo {pages} {074502} (\bibinfo {year}
  {2020})}\BibitemShut {NoStop}%
\bibitem [{\citenamefont {Adamczyk}\ \emph {et~al.}(2017)\citenamefont
  {Adamczyk} \emph {et~al.}}]{Adamczyk:2017iwn}%
  \BibitemOpen
  \bibfield  {author} {\bibinfo {author} {\bibfnamefont {L.}~\bibnamefont
  {Adamczyk}} \emph {et~al.} (\bibinfo {collaboration} {STAR collaboration}),\
  }\href {\doibase 10.1103/PhysRevC.96.044904} {\bibfield  {journal} {\bibinfo
  {journal} {Phys.\ Rev.\ C}\ }\textbf {\bibinfo {volume} {96}},\ \bibinfo
  {pages} {044904} (\bibinfo {year} {2017})}\BibitemShut {NoStop}%
\bibitem [{\citenamefont {Luo}\ and\ \citenamefont {Xu}(2017)}]{Luo:2017faz}%
  \BibitemOpen
  \bibfield  {author} {\bibinfo {author} {\bibfnamefont {X.}~\bibnamefont
  {Luo}}\ and\ \bibinfo {author} {\bibfnamefont {N.}~\bibnamefont {Xu}},\
  }\href {\doibase 10.1007/s41365-017-0257-0} {\bibfield  {journal} {\bibinfo
  {journal} {Nucl. Sci. Tech.}\ }\textbf {\bibinfo {volume} {28}},\ \bibinfo
  {pages} {112} (\bibinfo {year} {2017})}\BibitemShut {NoStop}%
\bibitem [{\citenamefont {Garg}\ \emph {et~al.}(2013)\citenamefont {Garg},
  \citenamefont {Mishra}, \citenamefont {Netrakanti}, \citenamefont {Mohanty},
  \citenamefont {Mohanty}, \citenamefont {Singh},\ and\ \citenamefont
  {Xu}}]{Garg:2013ata}%
  \BibitemOpen
  \bibfield  {author} {\bibinfo {author} {\bibfnamefont {P.}~\bibnamefont
  {Garg}}, \bibinfo {author} {\bibfnamefont {D.~K.}\ \bibnamefont {Mishra}},
  \bibinfo {author} {\bibfnamefont {P.~K.}\ \bibnamefont {Netrakanti}},
  \bibinfo {author} {\bibfnamefont {B.}~\bibnamefont {Mohanty}}, \bibinfo
  {author} {\bibfnamefont {A.~K.}\ \bibnamefont {Mohanty}}, \bibinfo {author}
  {\bibfnamefont {B.~K.}\ \bibnamefont {Singh}}, \ and\ \bibinfo {author}
  {\bibfnamefont {N.}~\bibnamefont {Xu}},\ }\href {\doibase
  10.1016/j.physletb.2013.09.019} {\bibfield  {journal} {\bibinfo  {journal}
  {Phys. Lett. B}\ }\textbf {\bibinfo {volume} {726}},\ \bibinfo {pages} {691}
  (\bibinfo {year} {2013})}\BibitemShut {NoStop}%
\bibitem [{\citenamefont {Asakawa}\ and\ \citenamefont
  {Kitazawa}(2016)}]{Asakawa:2015ybt}%
  \BibitemOpen
  \bibfield  {author} {\bibinfo {author} {\bibfnamefont {M.}~\bibnamefont
  {Asakawa}}\ and\ \bibinfo {author} {\bibfnamefont {M.}~\bibnamefont
  {Kitazawa}},\ }\href {\doibase 10.1016/j.ppnp.2016.04.002} {\bibfield
  {journal} {\bibinfo  {journal} {Prog. Part. Nucl. Phys.}\ }\textbf {\bibinfo
  {volume} {90}},\ \bibinfo {pages} {299} (\bibinfo {year} {2016})}\BibitemShut
  {NoStop}%
\bibitem [{\citenamefont {Luo}(2015)}]{eff_xiaofeng}%
  \BibitemOpen
  \bibfield  {author} {\bibinfo {author} {\bibfnamefont {X.}~\bibnamefont
  {Luo}},\ }\href {\doibase 10.1103/PhysRevC.91.034907} {\bibfield  {journal}
  {\bibinfo  {journal} {Phys. Rev.}\ }\textbf {\bibinfo {volume} {C91}},\
  \bibinfo {pages} {034907} (\bibinfo {year} {2015})}\BibitemShut {NoStop}%
\bibitem [{\citenamefont {Skokov}\ \emph {et~al.}(2013)\citenamefont {Skokov},
  \citenamefont {Friman},\ and\ \citenamefont {Redlich}}]{Skokov:2012ds}%
  \BibitemOpen
  \bibfield  {author} {\bibinfo {author} {\bibfnamefont {V.}~\bibnamefont
  {Skokov}}, \bibinfo {author} {\bibfnamefont {B.}~\bibnamefont {Friman}}, \
  and\ \bibinfo {author} {\bibfnamefont {K.}~\bibnamefont {Redlich}},\ }\href
  {\doibase 10.1103/PhysRevC.88.034911} {\bibfield  {journal} {\bibinfo
  {journal} {Phys. Rev.}\ }\textbf {\bibinfo {volume} {C88}},\ \bibinfo {pages}
  {034911} (\bibinfo {year} {2013})}\BibitemShut {NoStop}%
\bibitem [{\citenamefont {Sugiura}\ \emph {et~al.}(2019)\citenamefont
  {Sugiura}, \citenamefont {Nonaka},\ and\ \citenamefont
  {Esumi}}]{Sugiura:2019toh}%
  \BibitemOpen
  \bibfield  {author} {\bibinfo {author} {\bibfnamefont {T.}~\bibnamefont
  {Sugiura}}, \bibinfo {author} {\bibfnamefont {T.}~\bibnamefont {Nonaka}}, \
  and\ \bibinfo {author} {\bibfnamefont {S.}~\bibnamefont {Esumi}},\ }\href
  {\doibase 10.1103/PhysRevC.100.044904} {\bibfield  {journal} {\bibinfo
  {journal} {Phys. Rev.}\ }\textbf {\bibinfo {volume} {C100}},\ \bibinfo
  {pages} {044904} (\bibinfo {year} {2019})}\BibitemShut {NoStop}%
\bibitem [{\citenamefont {Pandav}\ \emph {et~al.}(2019)\citenamefont {Pandav},
  \citenamefont {Mallick},\ and\ \citenamefont {Mohanty}}]{Pandav:2018bdx}%
  \BibitemOpen
  \bibfield  {author} {\bibinfo {author} {\bibfnamefont {A.}~\bibnamefont
  {Pandav}}, \bibinfo {author} {\bibfnamefont {D.}~\bibnamefont {Mallick}}, \
  and\ \bibinfo {author} {\bibfnamefont {B.}~\bibnamefont {Mohanty}},\ }\href
  {\doibase 10.1016/j.nuclphysa.2019.08.002} {\bibfield  {journal} {\bibinfo
  {journal} {Nucl. Phys. A}\ }\textbf {\bibinfo {volume} {991}},\ \bibinfo
  {pages} {121608} (\bibinfo {year} {2019})}\BibitemShut {NoStop}%
\bibitem [{\citenamefont {Kitazawa}\ and\ \citenamefont
  {Asakawa}(2012)}]{eff_kitazawa}%
  \BibitemOpen
  \bibfield  {author} {\bibinfo {author} {\bibfnamefont {M.}~\bibnamefont
  {Kitazawa}}\ and\ \bibinfo {author} {\bibfnamefont {M.}~\bibnamefont
  {Asakawa}},\ }\href {\doibase 10.1103/PhysRevC.86.024904,
  10.1103/PhysRevC.86.069902} {\bibfield  {journal} {\bibinfo  {journal} {Phys.
  Rev.}\ }\textbf {\bibinfo {volume} {C86}},\ \bibinfo {pages} {024904}
  (\bibinfo {year} {2012})},\ \bibinfo {note} {[Erratum: Phys.
  Rev.C86,069902(2012)]}\BibitemShut {NoStop}%
\bibitem [{\citenamefont {Bzdak}\ and\ \citenamefont {Koch}(2012)}]{eff_koch}%
  \BibitemOpen
  \bibfield  {author} {\bibinfo {author} {\bibfnamefont {A.}~\bibnamefont
  {Bzdak}}\ and\ \bibinfo {author} {\bibfnamefont {V.}~\bibnamefont {Koch}},\
  }\href {\doibase 10.1103/PhysRevC.86.044904} {\bibfield  {journal} {\bibinfo
  {journal} {Phys. Rev.}\ }\textbf {\bibinfo {volume} {C86}},\ \bibinfo {pages}
  {044904} (\bibinfo {year} {2012})}\BibitemShut {NoStop}%
\bibitem [{\citenamefont {Nonaka}\ \emph {et~al.}(2016)\citenamefont {Nonaka},
  \citenamefont {Sugiura}, \citenamefont {Esumi}, \citenamefont {Masui},\ and\
  \citenamefont {Luo}}]{tsukuba_eff_separate}%
  \BibitemOpen
  \bibfield  {author} {\bibinfo {author} {\bibfnamefont {T.}~\bibnamefont
  {Nonaka}}, \bibinfo {author} {\bibfnamefont {T.}~\bibnamefont {Sugiura}},
  \bibinfo {author} {\bibfnamefont {S.}~\bibnamefont {Esumi}}, \bibinfo
  {author} {\bibfnamefont {H.}~\bibnamefont {Masui}}, \ and\ \bibinfo {author}
  {\bibfnamefont {X.}~\bibnamefont {Luo}},\ }\href {\doibase
  10.1103/PhysRevC.94.034909} {\bibfield  {journal} {\bibinfo  {journal} {Phys.
  Rev.}\ }\textbf {\bibinfo {volume} {C94}},\ \bibinfo {pages} {034909}
  (\bibinfo {year} {2016})}\BibitemShut {NoStop}%
\bibitem [{\citenamefont {Bzdak}\ and\ \citenamefont
  {Koch}(2015)}]{eff_psd_volker}%
  \BibitemOpen
  \bibfield  {author} {\bibinfo {author} {\bibfnamefont {A.}~\bibnamefont
  {Bzdak}}\ and\ \bibinfo {author} {\bibfnamefont {V.}~\bibnamefont {Koch}},\
  }\href {\doibase 10.1103/PhysRevC.91.027901} {\bibfield  {journal} {\bibinfo
  {journal} {Phys. Rev.}\ }\textbf {\bibinfo {volume} {C91}},\ \bibinfo {pages}
  {027901} (\bibinfo {year} {2015})}\BibitemShut {NoStop}%
\bibitem [{\citenamefont {Luo}\ and\ \citenamefont
  {Nonaka}(2019)}]{Luo:2018ofd}%
  \BibitemOpen
  \bibfield  {author} {\bibinfo {author} {\bibfnamefont {X.}~\bibnamefont
  {Luo}}\ and\ \bibinfo {author} {\bibfnamefont {T.}~\bibnamefont {Nonaka}},\
  }\href {\doibase 10.1103/PhysRevC.99.044917} {\bibfield  {journal} {\bibinfo
  {journal} {Phys. Rev.}\ }\textbf {\bibinfo {volume} {C99}},\ \bibinfo {pages}
  {044917} (\bibinfo {year} {2019})}\BibitemShut {NoStop}%
\bibitem [{\citenamefont {Kitazawa}(2016)}]{eff_psd_kitazawa}%
  \BibitemOpen
  \bibfield  {author} {\bibinfo {author} {\bibfnamefont {M.}~\bibnamefont
  {Kitazawa}},\ }\href {\doibase 10.1103/PhysRevC.93.044911} {\bibfield
  {journal} {\bibinfo  {journal} {Phys. Rev.}\ }\textbf {\bibinfo {volume}
  {C93}},\ \bibinfo {pages} {044911} (\bibinfo {year} {2016})}\BibitemShut
  {NoStop}%
\bibitem [{\citenamefont {Nonaka}\ \emph {et~al.}(2017)\citenamefont {Nonaka},
  \citenamefont {Kitazawa},\ and\ \citenamefont {Esumi}}]{Nonaka:2017kko}%
  \BibitemOpen
  \bibfield  {author} {\bibinfo {author} {\bibfnamefont {T.}~\bibnamefont
  {Nonaka}}, \bibinfo {author} {\bibfnamefont {M.}~\bibnamefont {Kitazawa}}, \
  and\ \bibinfo {author} {\bibfnamefont {S.}~\bibnamefont {Esumi}},\ }\href
  {\doibase 10.1103/PhysRevC.95.064912} {\bibfield  {journal} {\bibinfo
  {journal} {Phys. Rev.}\ }\textbf {\bibinfo {volume} {C95}},\ \bibinfo {pages}
  {064912} (\bibinfo {year} {2017})}\BibitemShut {NoStop}%
\bibitem [{\citenamefont {Bzdak}\ \emph {et~al.}(2016)\citenamefont {Bzdak},
  \citenamefont {Holzmann},\ and\ \citenamefont {Koch}}]{binomial_breaking}%
  \BibitemOpen
  \bibfield  {author} {\bibinfo {author} {\bibfnamefont {A.}~\bibnamefont
  {Bzdak}}, \bibinfo {author} {\bibfnamefont {R.}~\bibnamefont {Holzmann}}, \
  and\ \bibinfo {author} {\bibfnamefont {V.}~\bibnamefont {Koch}},\ }\href
  {\doibase 10.1103/PhysRevC.94.064907} {\bibfield  {journal} {\bibinfo
  {journal} {Phys. Rev.}\ }\textbf {\bibinfo {volume} {C94}},\ \bibinfo {pages}
  {064907} (\bibinfo {year} {2016})}\BibitemShut {NoStop}%
\bibitem [{\citenamefont {Esumi}\ \emph {et~al.}(2021)\citenamefont {Esumi},
  \citenamefont {Nakagawa},\ and\ \citenamefont {Nonaka}}]{Esumi:2020xdo}%
  \BibitemOpen
  \bibfield  {author} {\bibinfo {author} {\bibfnamefont {S.}~\bibnamefont
  {Esumi}}, \bibinfo {author} {\bibfnamefont {K.}~\bibnamefont {Nakagawa}}, \
  and\ \bibinfo {author} {\bibfnamefont {T.}~\bibnamefont {Nonaka}},\ }\href
  {\doibase 10.1016/j.nima.2020.164802} {\bibfield  {journal} {\bibinfo
  {journal} {Nucl. Instrum. Meth. A}\ }\textbf {\bibinfo {volume} {987}},\
  \bibinfo {pages} {164802} (\bibinfo {year} {2021})}\BibitemShut {NoStop}%
\bibitem [{\citenamefont {Nonaka}\ \emph {et~al.}(2018)\citenamefont {Nonaka},
  \citenamefont {Kitazawa},\ and\ \citenamefont {Esumi}}]{Nonaka:2018mgw}%
  \BibitemOpen
  \bibfield  {author} {\bibinfo {author} {\bibfnamefont {T.}~\bibnamefont
  {Nonaka}}, \bibinfo {author} {\bibfnamefont {M.}~\bibnamefont {Kitazawa}}, \
  and\ \bibinfo {author} {\bibfnamefont {S.}~\bibnamefont {Esumi}},\ }\href
  {\doibase 10.1016/j.nima.2018.08.013} {\bibfield  {journal} {\bibinfo
  {journal} {Nucl. Instrum. Meth.}\ }\textbf {\bibinfo {volume} {A906}},\
  \bibinfo {pages} {10} (\bibinfo {year} {2018})}\BibitemShut {NoStop}%
\bibitem [{\citenamefont {Barlow}(2002)}]{Barlow:2002yb}%
  \BibitemOpen
  \bibfield  {author} {\bibinfo {author} {\bibfnamefont {R.}~\bibnamefont
  {Barlow}},\ }in\ \href
  {http://www.ippp.dur.ac.uk/Workshops/02/statistics/proceedings//barlow.pdf}
  {\emph {\bibinfo {booktitle} {{Advanced Statistical Techniques in Particle
  Physics. Proceedings, Conference, Durham, UK, March 18-22, 2002}}}}\
  (\bibinfo {year} {2002})\ pp.\ \bibinfo {pages} {134--144}\BibitemShut
  {NoStop}%
\bibitem [{\citenamefont {Bleicher}\ \emph {et~al.}(1999)\citenamefont
  {Bleicher} \emph {et~al.}}]{Bleicher:1999xi}%
  \BibitemOpen
  \bibfield  {author} {\bibinfo {author} {\bibfnamefont {M.}~\bibnamefont
  {Bleicher}} \emph {et~al.},\ }\href {\doibase 10.1088/0954-3899/25/9/308}
  {\bibfield  {journal} {\bibinfo  {journal} {J.\ Phys.\ G}\ }\textbf {\bibinfo
  {volume} {25}},\ \bibinfo {pages} {1859} (\bibinfo {year}
  {1999})}\BibitemShut {NoStop}%
\bibitem [{\citenamefont {Ohnishi}\ \emph {et~al.}(2016)\citenamefont
  {Ohnishi}, \citenamefont {Kitazawa},\ and\ \citenamefont
  {Asakawa}}]{Ohnishi:2016bdf}%
  \BibitemOpen
  \bibfield  {author} {\bibinfo {author} {\bibfnamefont {Y.}~\bibnamefont
  {Ohnishi}}, \bibinfo {author} {\bibfnamefont {M.}~\bibnamefont {Kitazawa}}, \
  and\ \bibinfo {author} {\bibfnamefont {M.}~\bibnamefont {Asakawa}},\ }\href
  {\doibase 10.1103/PhysRevC.94.044905} {\bibfield  {journal} {\bibinfo
  {journal} {Phys. Rev. C}\ }\textbf {\bibinfo {volume} {94}},\ \bibinfo
  {pages} {044905} (\bibinfo {year} {2016})}\BibitemShut {NoStop}%
\bibitem [{\citenamefont {Bzdak}\ \emph {et~al.}(2013)\citenamefont {Bzdak},
  \citenamefont {Koch},\ and\ \citenamefont {Skokov}}]{Bzdak:2012an}%
  \BibitemOpen
  \bibfield  {author} {\bibinfo {author} {\bibfnamefont {A.}~\bibnamefont
  {Bzdak}}, \bibinfo {author} {\bibfnamefont {V.}~\bibnamefont {Koch}}, \ and\
  \bibinfo {author} {\bibfnamefont {V.}~\bibnamefont {Skokov}},\ }\href
  {\doibase 10.1103/PhysRevC.87.014901} {\bibfield  {journal} {\bibinfo
  {journal} {Phys.\ Rev.\ C}\ }\textbf {\bibinfo {volume} {87}},\ \bibinfo
  {pages} {014901} (\bibinfo {year} {2013})}\BibitemShut {NoStop}%
\bibitem [{\citenamefont {Braun-Munzinger}\ \emph {et~al.}(2017)\citenamefont
  {Braun-Munzinger}, \citenamefont {Rustamov},\ and\ \citenamefont
  {Stachel}}]{Braun-Munzinger:2016yjz}%
  \BibitemOpen
  \bibfield  {author} {\bibinfo {author} {\bibfnamefont {P.}~\bibnamefont
  {Braun-Munzinger}}, \bibinfo {author} {\bibfnamefont {A.}~\bibnamefont
  {Rustamov}}, \ and\ \bibinfo {author} {\bibfnamefont {J.}~\bibnamefont
  {Stachel}},\ }\href {\doibase 10.1016/j.nuclphysa.2017.01.011} {\bibfield
  {journal} {\bibinfo  {journal} {Nucl. Phys. A}\ }\textbf {\bibinfo {volume}
  {960}},\ \bibinfo {pages} {114} (\bibinfo {year} {2017})}\BibitemShut
  {NoStop}%
\bibitem [{\citenamefont {Vovchenko}\ and\ \citenamefont
  {Koch}(2021)}]{Vovchenko:2020kwg}%
  \BibitemOpen
  \bibfield  {author} {\bibinfo {author} {\bibfnamefont {V.}~\bibnamefont
  {Vovchenko}}\ and\ \bibinfo {author} {\bibfnamefont {V.}~\bibnamefont
  {Koch}},\ }\href {\doibase 10.1103/PhysRevC.103.044903} {\bibfield  {journal}
  {\bibinfo  {journal} {Phys. Rev. C}\ }\textbf {\bibinfo {volume} {103}},\
  \bibinfo {pages} {044903} (\bibinfo {year} {2021})}\BibitemShut {NoStop}%
\bibitem [{\citenamefont {Vovchenko}\ \emph {et~al.}(2021)\citenamefont
  {Vovchenko}, \citenamefont {Koch},\ and\ \citenamefont
  {Shen}}]{Vovchenko:2021kxx}%
  \BibitemOpen
  \bibfield  {author} {\bibinfo {author} {\bibfnamefont {V.}~\bibnamefont
  {Vovchenko}}, \bibinfo {author} {\bibfnamefont {V.}~\bibnamefont {Koch}}, \
  and\ \bibinfo {author} {\bibfnamefont {C.}~\bibnamefont {Shen}},\ }\href@noop
  {} {\  (\bibinfo {year} {2021})},\ \Eprint {http://arxiv.org/abs/2107.00163}
  {arXiv:2107.00163 [hep-ph]} \BibitemShut {NoStop}%
\bibitem [{\citenamefont {Braun-Munzinger}\ \emph {et~al.}(2021)\citenamefont
  {Braun-Munzinger}, \citenamefont {Friman}, \citenamefont {Redlich},
  \citenamefont {Rustamov},\ and\ \citenamefont
  {Stachel}}]{Braun-Munzinger:2020jbk}%
  \BibitemOpen
  \bibfield  {author} {\bibinfo {author} {\bibfnamefont {P.}~\bibnamefont
  {Braun-Munzinger}}, \bibinfo {author} {\bibfnamefont {B.}~\bibnamefont
  {Friman}}, \bibinfo {author} {\bibfnamefont {K.}~\bibnamefont {Redlich}},
  \bibinfo {author} {\bibfnamefont {A.}~\bibnamefont {Rustamov}}, \ and\
  \bibinfo {author} {\bibfnamefont {J.}~\bibnamefont {Stachel}},\ }\href
  {\doibase 10.1016/j.nuclphysa.2021.122141} {\bibfield  {journal} {\bibinfo
  {journal} {Nucl. Phys. A}\ }\textbf {\bibinfo {volume} {1008}},\ \bibinfo
  {pages} {122141} (\bibinfo {year} {2021})}\BibitemShut {NoStop}%
\bibitem [{\citenamefont {Mishra}\ and\ \citenamefont
  {Garg}(2017)}]{Mishra:2017mor}%
  \BibitemOpen
  \bibfield  {author} {\bibinfo {author} {\bibfnamefont {D.~K.}\ \bibnamefont
  {Mishra}}\ and\ \bibinfo {author} {\bibfnamefont {P.}~\bibnamefont {Garg}},\
  }\href@noop {} {\  (\bibinfo {year} {2017})},\ \Eprint
  {http://arxiv.org/abs/1706.04012} {arXiv:1706.04012 [nucl-th]} \BibitemShut
  {NoStop}%
\end{thebibliography}%
\appendix
\section*{Supplemental Material}
\subsection{Cumulants and efficiency correction}
The shape of the distribution is characterized by $r$th-order 
moments $\ave{N^{r}}$ which are defined as
\begin{equation}
	\mu_{r} = \frac{\partial^{r}}{\partial\theta^{r}}G(\theta)|_{\theta=0},\;\; 
	G(\theta) = \sum_{N}e^{N\theta}P(N), 
\end{equation}
where $G(\theta)$ is the moment generating function, 
and $P(N)$ is the probability distribution function to find 
$N$ particles~\cite{Asakawa:2015ybt}. 
Similarly, the $r$th-order cumulant, $C_{r}$, is defined as 
\begin{equation}
	C_{r} = \frac{\partial^{r}}{\partial\theta^{r}}K(\theta)|_{\theta=0},\;\; 
	K(\theta) = {\rm ln}\sum_{N}e^{N\theta}P(N), 
\end{equation}
where $K(\theta)$ is the cumulant generating function. 
The $r$th-order cumulant is then calculated recursively from lower-order moments as
\begin{equation}
	C_{r} = \mu_{r} - \sum_{m=1}^{r-1}\binom{r-1}{m-1}C_{m}\mu_{r-m}. 
\end{equation}

In experiments, the measured cumulants have to be corrected for detector efficiencies. 
The efficiency correction of the $r$th-order cumulant, $C_{r}^{cor}$ is performed with the 
following formulas~\cite{Nonaka:2017kko,Luo:2018ofd}:
\begin{widetext}
\begin{eqnarray}
	C^{cor} &=& \ave{q_{(1,1)}}_{\rm c}, 
        \label{eq:mk_1} \\ \nonumber \\
	C_{2}^{cor} &=& \ave{q_{(1,1)}^{2}}_{\rm c} + \ave{q_{(2,1)}}_{\rm c} - \ave{q_{(2,2)}}_{\rm c}, 
        \label{eq:mk_2} \\ \nonumber \\
	C_{3}^{cor}
        &=& \ave{q_{(1,1)}^{3}}_{\rm c} 
        + 3\ave{q_{(1,1)}q_{(2,1)}}_{\rm c} - 3\ave{q_{(1,1)}q_{(2,2)}}_{\rm c}
        + \ave{q_{(3,1)}}_{\rm c} - 3\ave{q_{(3,2)}}_{\rm c}
        + 2\ave{q_{(3,3)}}_{\rm c}, \label{eq:mk_3}
        \\ \nonumber \\
	C_{4}^{cor}
        &=& \ave{q_{(1,1)}^{4}}_{\rm c} 
        + 6\ave{q_{(1,1)}^{2}q_{(2,1)}}_{\rm c} - 6\ave{q_{(1,1)}^{2}q_{(2,2)}}_{\rm c} + 4\ave{q_{(1,1)}q_{(3,1)}}_{\rm c} + 3\ave{q_{(2,1)}^{2}}_{\rm c}
	\nonumber \\
	&& + 3\ave{q_{(2,2)}^{2}}_{\rm c} - 12\ave{q_{(1,1)}q_{(3,2)}}_{\rm c} + 8\ave{q_{(1,1)}q_{(3,3)}}_{\rm c} -  6\ave{q_{(2,1)}q_{(2,2)}}_{\rm c}
        \nonumber \\
	&& + \ave{q_{(4,1)}}_{\rm c} - 7\ave{q_{(4,2)}}_{\rm c} 
        + 12\ave{q_{(4,3)}}_{\rm c} - 6\ave{q_{(4,4)}}_{\rm c}, 
	\label{eq:mk_4} \\ \nonumber \\
	C_{5}^{cor} &=& \ave{q_{(1,1)}^{5}}_{\rm c} 
				+ 10\ave{q_{(1,1)}^{3}q_{(2,1)}}_{\rm c} - 10\ave{q_{(1,1)}^{3}q_{(2,2)}}_{\rm c} + 10\ave{q_{(1,1)}^{2}q_{(3,1)}}_{\rm c} - 30\ave{q_{(1,1)}^{2}q_{(3,2)}}_{\rm c}  \nonumber \\
			     && + 20\ave{q_{(1,1)}^{2}q_{(3,3)}}_{\rm c} + 15\ave{q_{(2,2)}^{2}q_{(1,1)}}_{\rm c} + 15\ave{q_{(2,1)}^{2}q_{(1,1)}}_{\rm c} - 30\ave{q_{(1,1)}q_{(2,1)}q_{(2,2)}}_{\rm c} \nonumber \\ 
			     && +  5\ave{q_{(1,1)}q_{(4,1)}}_{\rm c} - 35\ave{q_{(1,1)}q_{(4,2)}}_{\rm c} + 60\ave{q_{(1,1)}q_{(4,3)}}_{\rm c} - 30\ave{q_{(1,1)}q_{(4,4)}}_{\rm c}  \nonumber \\
			     && + 10\ave{q_{(2,1)}q_{(3,1)}}_{\rm c} - 30\ave{q_{(2,1)}q_{(3,2)}}_{\rm c} + 20\ave{q_{(2,1)}q_{(3,3)}}_{\rm c}    \nonumber \\ 
			     && - 10\ave{q_{(2,2)}q_{(3,1)}}_{\rm c} + 30\ave{q_{(2,2)}q_{(3,2)}}_{\rm c} - 20\ave{q_{(2,2)}q_{(3,3)}}_{\rm c}   \nonumber \\
			     && + \ave{q_{(5,1)}}_{\rm c} - 15\ave{q_{(5,2)}}_{\rm c} + 50\ave{q_{(5,3)}}_{\rm c} - 60\ave{q_{(5,4)}}_{\rm c} + 24\ave{q_{(5,5)}}_{\rm c}, \label{eq:mk_5} \\ \nonumber \\
	C_{6}^{cor} &=& \ave{q_{(1,1)}^{6}}_{\rm c} + 15\ave{q_{(1,1)}^{4}q_{(2,1)}}_{\rm c} - 15\ave{q_{(1,1)}^{4}q_{(2,2)}}_{\rm c} + 20\ave{q_{(1,1)}^{3}q_{(3,1)}}_{\rm c} - 60\ave{q_{(1,1)}^{3}q_{(3,2)}}_{\rm c}  \nonumber \\ 
			     && + 40\ave{q_{(1,1)}^{3}q_{(3,3)}}_{\rm c} - 90\ave{q_{(1,1)}^{2}q_{(2,2)}q_{(2,1)}}_{\rm c} + 45\ave{q_{(1,1)}^{2}q_{(2,1)}^{2}}_{\rm c} + 45\ave{q_{(1,1)}^{2}q_{(2,2)}^{2}}_{\rm c} \nonumber \\ 
			     && + 15\ave{q_{(2,1)}^{3}}_{\rm c} - 15\ave{q_{(2,2)}^{3}}_{\rm c} + 15\ave{q_{(1,1)}^{2}q_{(4,1)}}_{\rm c} - 105\ave{q_{(1,1)}^{2}q_{(4,2)}}_{\rm c} + 180\ave{q_{(1,1)}^{2}q_{(4,3)}}_{\rm c} - 90\ave{q_{(1,1)}^{2}q_{(4,4)}}_{\rm c} \nonumber \\ 
			     && - 45\ave{q_{(2,1)}^{2}q_{(2,2)}}_{\rm c} + 45\ave{q_{(2,2)}^{2}q_{(2,1)}}_{\rm c} + 60\ave{q_{(1,1)}q_{(2,1)}q_{(3,1)}}_{\rm c} - 180\ave{q_{(1,1)}q_{(2,1)}q_{(3,2)}}_{\rm c}    \nonumber \\ 
			     && + 120\ave{q_{(1,1)}q_{(2,1)}q_{(3,3)}}_{\rm c} - 60\ave{q_{(1,1)}q_{(2,2)}q_{(3,1)}}_{\rm c} + 180\ave{q_{(1,1)}q_{(2,2)}q_{(3,2)}}_{\rm c} - 120\ave{q_{(1,1)}q_{(2,2)}q_{(3,3)}}_{\rm c}  \nonumber \\
			     && + 6\ave{q_{(1,1)}q_{(5,1)}}_{\rm c} - 90\ave{q_{(1,1)}q_{(5,2)}}_{\rm c} + 300\ave{q_{(1,1)}q_{(5,3)}}_{\rm c} - 360\ave{q_{(1,1)}q_{(5,4)}}_{\rm c} + 144\ave{q_{(1,1)}q_{(5,5)}}_{\rm c} \nonumber \\
			     && + 15\ave{q_{(2,1)}q_{(4,1)}}_{\rm c} - 105\ave{q_{(2,1)}q_{(4,2)}}_{\rm c} + 180\ave{q_{(2,1)}q_{(4,3)}}_{\rm c} - 90\ave{q_{(2,1)}q_{(4,4)}}_{\rm c}   \nonumber \\
			     && - 15\ave{q_{(2,2)}q_{(4,1)}}_{\rm c} + 105\ave{q_{(2,2)}q_{(4,2)}}_{\rm c} - 180\ave{q_{(2,2)}q_{(4,3)}}_{\rm c} + 90\ave{q_{(2,2)}q_{(4,4)}}_{\rm c} \nonumber \\
			     && + 10\ave{q_{(3,1)}^{2}}_{\rm c} - 60\ave{q_{(3,1)}q_{(3,2)}}_{\rm c}+ 40\ave{q_{(3,1)}q_{(3,3)}}_{\rm c} + 90\ave{q_{(3,2)}^{2}}_{\rm c} - 120\ave{q_{(3,2)}q_{(3,3)}}_{\rm c} + 40\ave{q_{(3,3)}^{2}}_{\rm c}      \nonumber \\ 
			     && + \ave{q_{(6,1)}}_{\rm c} - 31\ave{q_{(6,2)}}_{\rm c} + 180\ave{q_{(6,3)}}_{\rm c} - 390\ave{q_{(6,4)}}_{\rm c} + 360\ave{q_{(6,5)}}_{\rm c}  - 120\ave{q_{(6,6)}}_{\rm c},
        \label{eq:mk_6} \nonumber \\ 
\end{eqnarray}
with
\begin{equation}
	q_{(r,s)} = q_{(a^r/p^s)}
        = \sum_{i=1}^{M} (a_{i}^{r}/p_{i}^{s}) n_{i}, 
        \label{eq:multi_q}
\end{equation}
\end{widetext}
where angle brackets represent represents a cumulant, $M$ represents the number of efficiency bins, 
$a_{i}$ and $p_{i}$ represent the electric charge and efficiency at the $i$th efficiency bin, 
and $r$ and $s$ are the powers of $a_{i}$ and $p_{i}$, respectively.

\subsection{Efficiency distributions\label{sec:effdist}}
In general, cumulants of net-particle distributions are corrected for 
detector efficiencies, and are assumed to follow binomial distributions~\cite{eff_kitazawa,eff_koch,eff_xiaofeng,Nonaka:2017kko}. 
However, it was pointed out that the correction method loses its validity once the binomial assumption 
is broken~\cite{binomial_breaking,Nonaka:2018mgw}. 
In order to check the efficiency distributions of the STAR detector, 
Monte-Carlo simulations are employed in the STAR detector environment. 
The number of reconstructed protons is used to study non-binomial 
effect of the detector in the most severe environment 
in terms of the particle multiplicity, MC tracks are embedded  
in the data for Au+Au 0-5\% central collisions at $\sqrt{s_{\rm NN}}=$~200~GeV. 
Efficiencies are also studied by varying the embedded number of protons and antiprotons to see 
the dependence of non-binomial effects on the track density. 

\begin{figure*}[htbp]
	\begin{center}
	\includegraphics[width=140mm]{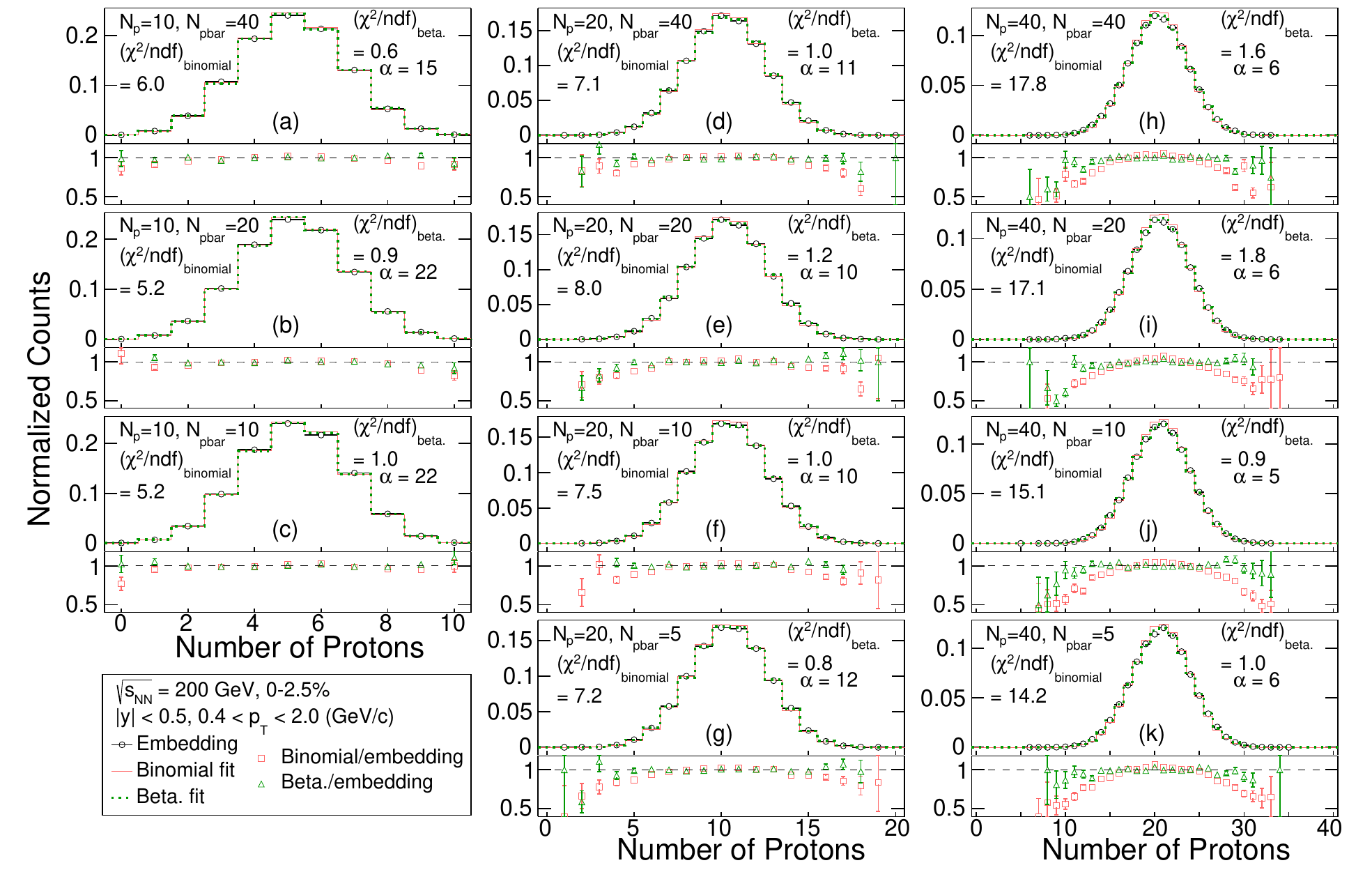}
	\end{center}
	\caption{
		Distributions of reconstructed protons (black circles) from 
		embedding simulations in 200~GeV Au+Au collisions at 0-2.5\% centrality. Red lines are 
		fits with the binomial distribution, and green dotted lines represent the 
		fit with the beta-binomial distributions using $\alpha$ that gives the minimal $\chi^{2}/{\rm ndf}$. 
		Each panel shows the result from the given combinations of embedded 
		protons and antiprotons. The ratio of fits to the embedding data is shown for each panel.
		}
	\label{fig:EmbeddingFitFinal}
\end{figure*}

Figure~\ref{fig:EmbeddingFitFinal} shows reconstructed protons as black circles. 
Each panel represents a different number of embedded protons and antiprotons. 
Embedding data is then fitted by the binomial distribution 
as shown by red solid lines in Fig.~\ref{fig:EmbeddingFitFinal}. 
The $\chi^{2}/{\rm ndf}$ values can be found in each panel. 
The ratio of the fitted function to the embedding data is shown in the lower panels.
It is found that the tail of the distributions in panel (e), (f),... (k) in Fig.~\ref{fig:EmbeddingFitFinal} 
deviates from binomial distributions. 
The beta-binomial distribution is employed to fit the embedding data in addition to the binomial distribution. 
The beta-binomial distribution is defined as 
\begin{equation}
	\beta(n:N,a,b) = \int_{0}^{1}d\varepsilon{\cal B}(\varepsilon,a,b)B(n;\varepsilon,N),
\end{equation}
with the beta distribution being 
\begin{equation}
	{\cal B}(\varepsilon;a,b) = \varepsilon^{a}(1-\varepsilon)^{b}/B(a,b),
\end{equation}
where $B(a,b)$ is the beta function. 
Each variable is defined through a model as follows.
It is known that the beta-binomial distribution is given by an urn model. 
Let us consider $N_{w}$ white balls and $N_{b}$ black balls in the urn. 
One draws a ball from the urn. If it is white (black), 
return two white (black) balls to the urn. 
This procedure is repeated $N$ times, then the resulting distribution 
of $n$ (white balls) is given by the beta-binomial distributions as $\beta(n;N,N_{w},N_{b})$. 
This is equivalent to $\beta(n;N,\alpha,\varepsilon)$, 
where $N_{w}=\alpha N$ with $\varepsilon=N_{w}/(N_{w}+N_{b})$.
Smaller $\alpha$ gives a broader distribution than the binomial, 
while the distribution becomes close to the binomial distribution with larger $\alpha$.
The beta-binomial distributions are numerically generated using various values of $\alpha$. 
They are compared to the embedding data to determine the best parameter $\alpha$ to 
fit the efficiency distributions. 

The fitted function is shown by the green dotted lines in Fig.~\ref{fig:EmbeddingFitFinal}. 
The values of $\chi^{2}/{\rm ndf}$ are found to be around unity in most of $(N_{p},N_{\bar p})$.
This indicates that the STAR efficiency follows the beta-binomial distribution 
in Au+Au collisions 0-2.5\% centrality at $\sqrt{s_{\rm NN}}=$~200~GeV. 
Similar checks are done for 2.5-5.0\% centrality. It is found that the efficiency distribution 
follows the beta-binomial distribution also for 2.5-5.0\% centrality.

\begin{table}[htbp]
	\begin{center}
	\begin{tabular}{ccc}\hline
		\multicolumn{1}{c}{\multirow{2}{*}{($N_{p}$,$N_{\bar p}$)}} & \multicolumn{2}{c}{$\chi^{2}/{\rm ndf}$ values} \\\cline{2-3}
		 &  binomial  &  beta-binomial ($\alpha$)  \\ \hline
		(10,10)   &  5.2  &  1.0 (22) \\ \hline
		(10,20)   &  5.2  &  0.9 (22) \\ \hline
		(10,40)   &  6.0  &  0.6 (15) \\ \hline
		(20,5)    &  7.2  &  0.8 (12) \\ \hline
		(20,10)   &  7.5  &  1.0 (10) \\ \hline
		(20,20)   &  8.0  &  1.2 (10) \\ \hline
		(20,40)   &  7.1  &  1.0 (11) \\ \hline
		(40,5)    &  14.2 &  1.0 (6) \\ \hline
		(40,10)   &  15.1 &  0.9 (5) \\ \hline
		(40,20)   &  17.1 &  1.8 (6) \\ \hline
		(40,40)   &  17.8 &  1.6 (6) \\ \hline
	\end{tabular}

	\end{center}
	\caption{ 
		$\chi^{2}/{\rm ndf}$ of fit results in Fig.~\ref{fig:EmbeddingFitFinal} 
		for the given combinations of embedded protons and antiprotons 
		from embedding simulations in 200~GeV Au+Au collisions at 0-2.5\% centrality.
		The non-binomial parameter $\alpha$ is also shown for results using the beta-binomial distribution. 
	}
	\label{tab:EmbeddingFitChi2}
\end{table}

\subsection{Non-binomial efficiency correction}
Two correction methods for non-binomial efficiency are employed. 
One is the unfolding approach to reconstruct the distribution itself~\cite{Esumi:2020xdo}, 
and the other one is the moment expansion to correct moments 
in an analytic way~\cite{Nonaka:2018mgw}.
Both methods utilize knowledge of the detector response between 
generated and reconstructed particles computed by detector simulations, 
which is called response matrix. 
From the embedding simulation discussed in Sec.~\ref{sec:effdist}, 
$\varepsilon$ and $\alpha$ are parameterized as a function 
of $N_{p}$ and $N_{\bar p}$. Using the parametrization, the 
4-dimensional response matrices between generated/reconstructed coordinates for protons and antiprotons are 
generated using 1 billion events.
Another two response matrices are also generated using $\alpha-\sigma$ 
and $\alpha+\sigma$, where $\sigma$ is the statistical uncertainty on $\alpha$. 
Furthermore, the response matrices of the binomial distribution are generated 
with track density dependence of the efficiency implemented 
\footnote{The track density here represents event-by-event (anti)protons 
in the generated coordinate, which cannot be considered in the conventional 
efficiency correction method. Note that the dependence of efficiency on 
the total multiplicity is already taken into account in the centrality bin width correction (CBWC).}~\cite{binomial_breaking}. 
Those response matrices are utilized to correct for the detector effects.
We have checked that both the unfolding approach and the moment expansion method
~\cite{Esumi:2020xdo,Nonaka:2018mgw} give consistent cumulant results. 
Thus, we focus on the results of the unfolding approach in the rest of this section.

\begin{figure}[htbp]
	\begin{center}
	\includegraphics[width=85mm]{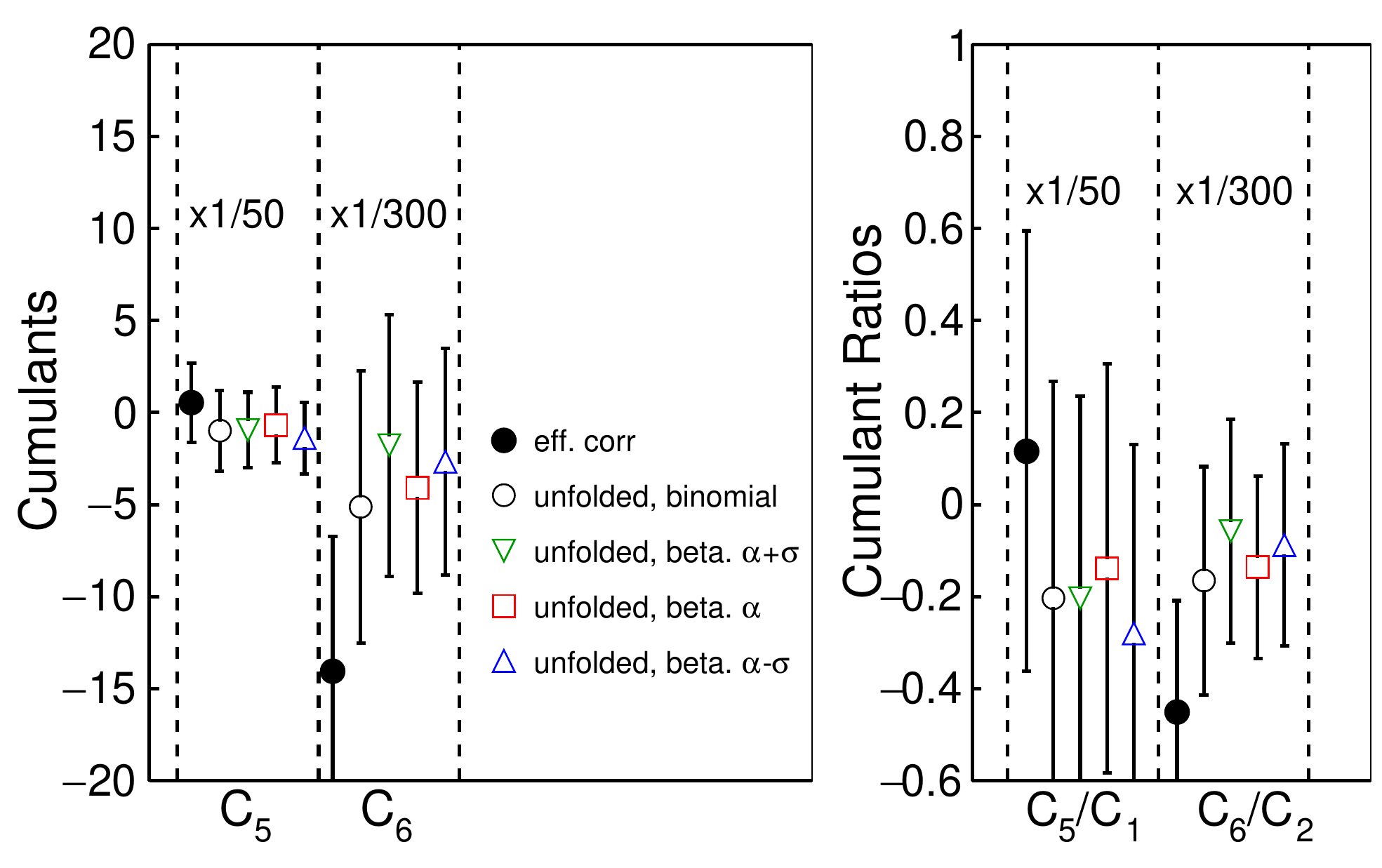}
	\end{center}
	\caption{
		Cumulants and their ratios up to the sixth order corrected for non-binomial efficiencies 
		for 200~GeV Au+Au collisions at 0-5\% centrality. The CBWC is applied for 
		2.5\% centrality bin width.
		Results from the conventional efficiency correction are shown as black filled circles, 
		results from the unfolding with the binomial detector response are shown as black open circles, 
		and results from beta-binomial detector response with $\alpha+\sigma$, $\alpha$ 
		and $\alpha-\sigma$ are shown in green triangles, red squares and blue triangles, respectively. 
		$C_{5}$, $C_{6}$, $C_{2}/C_{1}$, $C_{5}/C_{1}$ and $C_{6}/C_{2}$ are scaled by constant shown 
		in each column.
		}
	\label{fig:UnfFinal}
\end{figure}

It should be noted that in Ref.~\cite{Adam:2020unf} the effect of non-binomial effects of up to the fourth-order cumulants and their ratios are discussed and found to be small within statistical uncertainties.
Therefore, we present the effects on the higher-order cumulants and ratios, $C_{5}$, $C_{6}$, $C_{5}/C_{1}$, and $C_{6}/C_{2}$ 
for Au+Au collisions at $\sqrt{s_{\rm NN}}=$~200~GeV 0-5\% centrality as 
shown in Fig.~\ref{fig:UnfFinal}.  
For each column, the first two points are the results corrected for  
the binomial detector response, and last three points are from 
the unfolding with beta-binomial response using 
different non-binomial parameters. The dependence of efficiency 
on the track density ($N_{p}$ and $N_{\bar p}$ themselves) is taken into account for the last four points. 
Those results are arbitrarily ordered from left to right 
as response matrices have larger deviations compared 
to the binomial distribution.
They are calculated for 0-2.5\% and 2.5-5.0\% centralities 
separately and averaged to 0-5\% centrality. 
Table~\ref{tab:final} summarizes the cumulant ratios and their errors.
Results corrected for beta-binomial response matrices with non-binomial 
parameters being $\alpha\pm\sigma$ are considered as systematic 
uncertainties.
Statistical and systematic uncertainties are added in quadrature to calculate total uncertainties.
The deviations of results corrected for non-binomial efficiencies  
compared to the binomial efficiency correction is found to be 
less than $1$ sigma for $C_{5}/C_{1}$ and $C_{6}/C_{2}$. 

Again, we note that these results are from 200~GeV Au+Au collisions at 0-5\% centrality. 
This is the highest Au+Au collision energy at RHIC, 
where the high multiplicity is expected to result in the largest non-binomial detector effects.
It is noteworthy that the experimental data for 200~GeV has the 
largest statistics among the studied datasets. 
Therefore, we conclude that the non-binomial detector effects on higher-order 
cumulants and their ratios would be within errors for all BES-I energies. 

\begin{table*}[htbp]
	\begin{center}
	\begin{tabular}{cccc}\hline
		Cumulant ratio & binomial$\pm$stat.err  &  beta.$\pm$stat.err$\pm$sys.err  & significance \\ \hline
		$C_{2}/C_{1}$  &  $1.3\pm1.2\times10^{-3}$  &  $1.2\pm1.3\times10^{-3}\pm3.1\times10^{-2}$ & $3.1$  \\ \hline 
		$C_{3}/C_{2}$  &  $0.13\pm1.2\times10^{-2}$  &  $0.13\pm1.2\times10^{-2}\pm3.0\times10^{-3}$ & $4.7\times10^{-2}$  \\ \hline 
		$C_{4}/C_{2}$  &  $1.1\pm0.21$          &  $0.97\pm0.21\pm8.4\times10^{-2}$        & $4.2\times10^{-1}$  \\ \hline 
		$C_{5}/C_{1}$  &  $0.1\pm0.48$         &  $-0.14\pm0.44\pm0.11$              & $3.8\times10^{-1}$  \\ \hline 
		$C_{6}/C_{2}$  &  $-0.45\pm0.24$        &  $-0.14\pm0.20\pm6.5\times10^{-2}$       & $1.0$ \\ \hline 
	\end{tabular}
	\end{center}
	\caption{ 
	Cumulant ratios and their statistical errors (second column) from the conventional 
	efficiency correction with the binomial detector response, and 
	(third column) from unfolding with the beta-binomial detector response. 
	Systematic errors are also shown for the beta-binomial case.
	The last column shows the difference between the two results normalized by the total errors 
	mentioned above.
	}
	\label{tab:final}
\end{table*}

\subsection{Effect of centrality bin width correction (CBWC)}
\begin{figure}[htbp]
	\begin{center}
	\includegraphics[width=100mm]{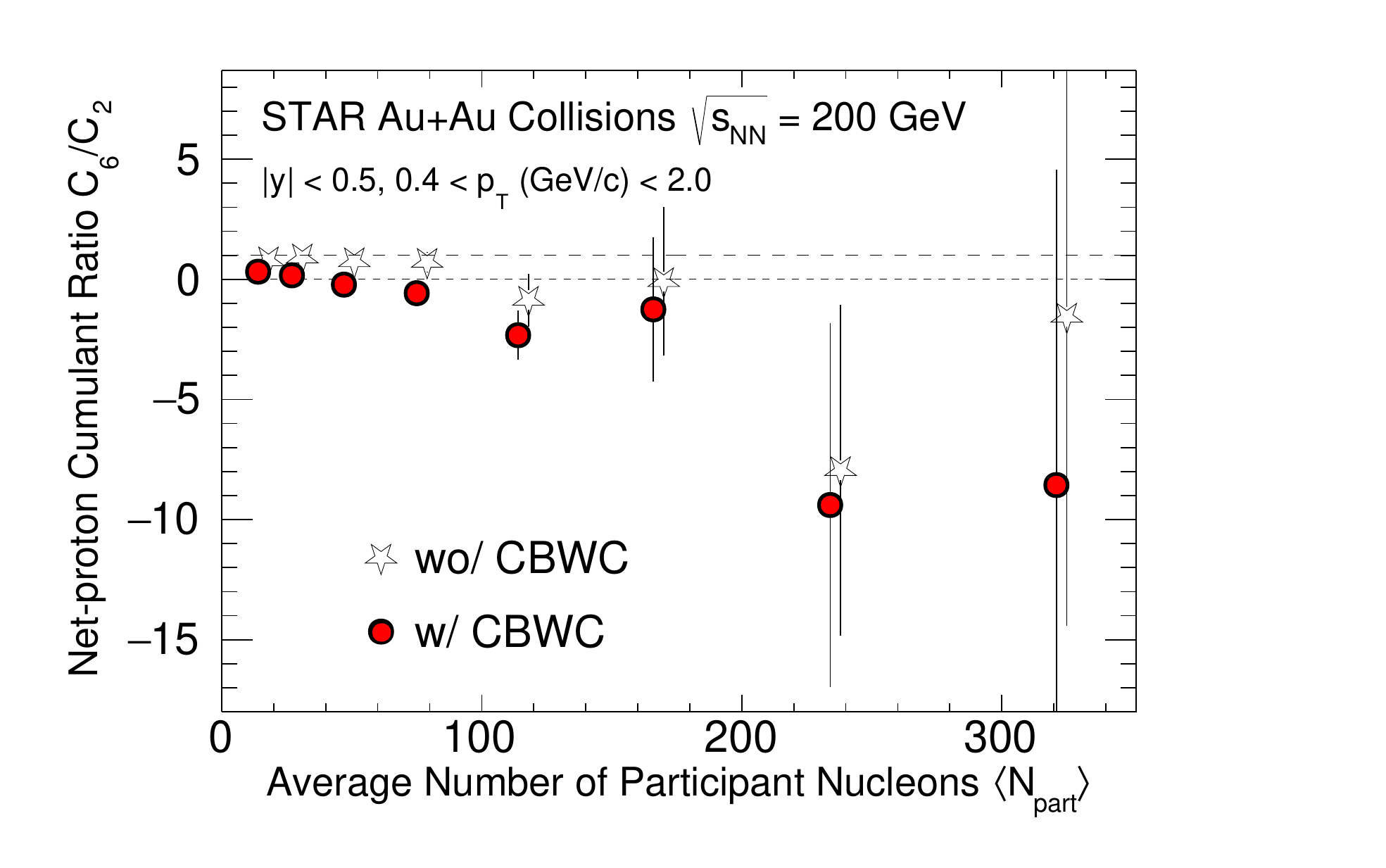}
	\end{center}
	\caption{
	Collision centrality dependence of net-proton 
	$C_{6}/C_{2}$ in Au+Au collisions for $\sqrt{s_{\rm NN}}=$~200~GeV 
	within $|y|<0.5$ and $0.4<p_{T}\;({\rm GeV/c})<2.0$. 
	Results with and without the CBWC are overlaid. 
	The results are corrected for detector efficiencies.
	Points for different calculation methods are staggered horizontally to improve clarity.
		}
	\label{fig:wocbwc_Supplemental}
\end{figure}

Figure~\ref{fig:wocbwc_Supplemental} shows centrality dependence of net-proton $C_{6}/C_{2}$ from Au+Au $\sqrt{s_{\rm NN}}=200$~GeV collisions.
The centrality bin width correction (CBWC)~\cite{Luo:2017faz} is applied for the red markers to supress the effect from initial volume fluctuations. 
No volume correction is applied for the star markers. 
One can see that the volume fluctuations enhance the value of net-proton $C_{6}/C_{2}$. 
Note that there is no perfect solution to remove the volume fluctuations, hence the data driven approach, CBWC, is employed in the measurements.

\subsection{UrQMD calculations}
\begin{figure}[H]
\vspace{2mm}
	\begin{center}
	\includegraphics[width=85mm]{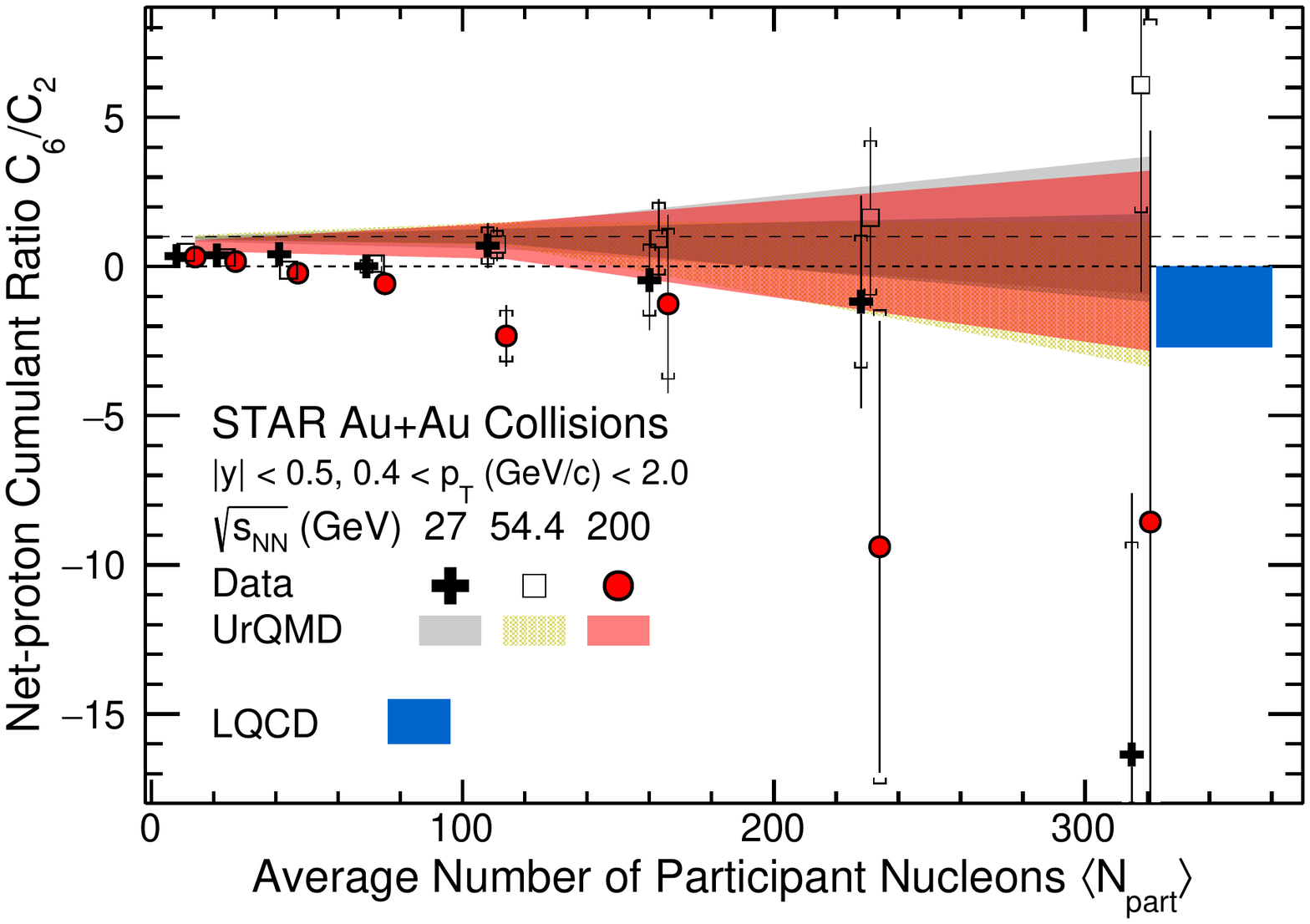}
	\end{center}
	\caption{
	Collision centrality dependence of net-proton 
	$C_{6}/C_{2}$ in Au+Au collisions for $\sqrt{s_{\rm NN}}=$~27, 54.4, and 200~GeV 
	within $|y|<0.5$ and $0.4<p_{T}\;({\rm GeV/c})<2.0$. 
	Points for different beam energies are staggered horizontally to improve clarity.
	Shaded and hatched bands show the results from UrQMD model calculations
	The lattice QCD calculations~\cite{Borsanyi:2018grb,Bazavov:2020bjn} 
	for $T=160$~MeV and $\mu_{\rm B}<$~110~MeV. 
	are shown as a blue band at $\ave{N_{\rm part}}\approx340$.
		}
	\label{fig:UrQMD_Supplemental}
\end{figure}

Figure~\ref{fig:UrQMD_Supplemental} shows centrality dependence of net-proton $C_{6}/C_{2}$ from 
experimental data and UrQMD calculations for $\sqrt{s_{\rm NN}}=$~27, 54.4, and 200~GeV. 
The UrQMD results are merged for 0-30\%, 30-60\%, and 60-80\% centralities to reduce statistical uncertainties. 
One can see that the $C_{6}/C_{2}$ values from three collision energies in UrQMD calculations are consistent with each other 
within uncertainties, and therefore they are merged to reduce statistical fluctuations in the paper. 

\end{document}